\documentclass[12pt]{article}

\usepackage{fullpage}
\usepackage{amsmath}
\usepackage{amssymb}
\usepackage{amsthm}
\usepackage[utf8]{inputenc}
\usepackage{tikz}
\usepackage{float}       
\usepackage{indentfirst} 
\usepackage{caption}
\usepackage[normalem]{ulem}

\usepackage{graphicx}  
\usepackage{epstopdf}
\usepackage{subfigure}

\usetikzlibrary{decorations.shapes}
\usetikzlibrary{shapes.geometric}

\theoremstyle{definition} 
\theoremstyle{plain} 
\theoremstyle{plain} 
\theoremstyle{plain} 
\theoremstyle{plain}

\newcommand{\cbb}{\mathbb{C}}
\newcommand{\ebb}{\mathbb{E}}
\newcommand{\nbb}{\mathbb{N}}

\newcommand{\rbb}{\mathbb{R}}

\newcommand{\zbb}{\mathbb{Z}}
\newcommand{\Ecal}{\mathcal{E}}

\def\E{{\mathcal{E}}}

\def\tg0{\tilde{g}_{\Ecal_0,T}}

\begin{document}
\pagenumbering{arabic}

\title{Transient Growth in Stochastic Burgers Flows}
\author{Diogo Po\c{c}as and Bartosz Protas\thanks{Corresponding author; Email {\tt bprotas@mcmaster.ca}}
\\ 
Department of Mathematics \& Statistics, \\
McMaster University  \\
Hamilton, Ontario L8S4K1, CANADA }

\date{\today}
\maketitle

\begin{abstract}
  This study considers the problem of the extreme behavior exhibited
  by solutions to Burgers equation subject to stochastic forcing. More
  specifically, we are interested in the maximum growth achieved by
  the ``enstrophy'' (the Sobolev $H^1$ seminorm of the solution) as a
  function of the initial enstrophy $\Ecal_0$, in particular, whether
  in the stochastic setting this growth is different than in the
  deterministic case considered by Ayala \& Protas (2011). This
  problem is motivated by questions about the effect of noise on the
  possible singularity formation in hydrodynamic models. The main
  quantities of interest in the stochastic problem are the expected
  value of the enstrophy and the enstrophy of the expected value of
  the solution. The stochastic Burgers equation is solved numerically
  with a Monte Carlo sampling approach. By studying solutions obtained
  for a range of optimal initial data and different noise magnitudes,
  we reveal different solution behaviors and it is demonstrated that
  the two quantities always bracket the enstrophy of the deterministic
  solution. The key finding is that the expected values of the
  enstrophy exhibit the same power-law dependence on the initial
  enstrophy $\Ecal_0$ as reported in the deterministic case.  This
  indicates that the stochastic excitation does not increase the
  extreme enstrophy growth beyond what is already observed in the
  deterministic case.
\end{abstract}

\begin{flushleft}
\end{flushleft}


\section{Introduction and Problem Statement}
\label{sec:intro}

Many open problems related to nonlinear partial differential equations
(PDEs) of mathematical physics concern the extreme behavior which can
be exhibited {by} their solutions. By this we mean, among other,
questions concerning the maximum possible growth of certain norms of
the solution of the PDE. From the physics point of view, these norms
measure different properties of the solution, such as generation of
small scales in the case of the Sobolev norms. The question of the
maximum possible growth of solution norms is also intrinsically linked
to the problem of existence of solutions to PDE problems in a given
functional space. More specifically, the loss of regularity of a
solution resulting from the formation of singularities usually
manifests itself in an unbounded growth of some solution norms in
finite time, typically referred to as ``blow-up''. While problems of
this type remain open for many important PDEs of mathematical physics,
most attention has been arguably given to establishing the regularity
of the three-dimensional (3D) Navier-Stokes equations \cite{d09}, a
problem which has been recognized by the Clay Mathematics Institute as
one of its ``millennium problems'' \cite{f00}. Analogous questions
also remain open for the 3D inviscid Euler equation \cite{gbk08}. The
problem we address in the present study is how the transient growth of
solutions to certain nonlinear PDEs is affected by the presence of
noise represented by a suitably defined stochastic forcing term in the
equation. More specifically, the key question is whether via some
interaction with the nonlinearity {and dissipation present in the
  system} such stochastic forcing may enhance or weaken the growth of
certain solution norms as compared to the deterministic case. In
particular, in the case of systems exhibiting finite-time blow-up in
the deterministic case it is interesting to know whether noise may
accelerate or delay the formation of a singularity, or perhaps even
prevent it entirely \cite{f15}.  These questions are of course nuanced
by the fact that they may be considered either for individual
trajectories or in suitable statistical terms. We add that transient
growth in linear stochastic systems is well understood
\cite{Kim2007arfm} and here we focus on the interaction of the
stochastic forcing with a particular type of nonlinearity.

Since this study is ultimately motivated by questions concerning
extreme behavior in hydrodynamic models, we will focus our attention
on the simplest model used in this context, namely, the
one-dimensional (1D) stochastic Burgers equation defined on a periodic
interval $[0,1]$
\begin{subequations}
\label{eq:burgers}
\begin{alignat}{2}
\partial_t u + {\frac{1}{2}\partial_x u^2} - \nu \partial_x^2 u &= \zeta & &\text{in} \ (0,T]\times(0,1), \label{eq:burgersA} \\
u(t,0)=u(t,1) \ \text{and} \ \partial_x u(t,0)& =\partial_x u(t,1) \qquad & & \text{for} \ t\in[0,T],\label{eq:burgersB} \\
u(0,x)&=g(x) &&\text{for} \ x\in(0,1), \label{eq:burgersC}
\end{alignat}
\end{subequations}
in which $T>0$ represents the length of the time window of interest,
$\nu > 0$ is the viscosity coefficient (hereafter we will use {$\nu =
  0.001$}) {and} $g \in H^1_p(0,1)$ is the initial condition,
where $H^1_p(0,1)$ denotes the Sobolev space of periodic functions
defined on $(0,1)$ with square integrable derivatives and the norm
given by \cite{af05}
\begin{equation}
\|u(t,\cdot)\|_{H_p^1}^2=\int_0^{1}|u(t,x)|^2+|\partial_x u(t,x)|^2 \, dx.
\label{eq:normH1}
\end{equation}
For simplicity, we will denote the time-space domain $D := (0,T]
\times (0,1)$ (``$:=$'' means ``equal to by definition''). In equation
\eqref{eq:burgersA} the stochastic forcing is given by a random field
$\zeta(t,x)$, $(t,x)\in D$. Therefore, at any point $(t,x)$ our
solution becomes a random variable $u={u(t,x;\omega)}$ for $\omega$ in
some probability space $\Omega$. We add that, while for other systems,
such as e.g., the Schr\"odinger equation \cite{ddm02a}, one may also
consider multiplicative noise, for {dissipative} models of the type
\eqref{eq:burgersA} one typically studies additive noise. The
reason is that, as argued in \cite[Section 5.5.2]{f15},
multiplicative noise tends to have effect similar to dissipative
terms, so if the equation already involves such a term, then no
major qualitative changes in the solution behavior can be expected.

A common approach to modelling stochastic excitation in PDE systems is
to describe it in terms of Gaussian noise white both in time and
space, and associated with an infinite-variance Wiener process.
However, as will be discussed in Section \ref{sec:noise}, such
a noise model does not ensure that individual solutions are
well defined in the Sobolev space $H^1_p$ and is therefore not
suitable for the problem considered here. Thus, for the remainder of
this paper, we shall restrict our attention to the case where $\zeta$
is the derivative of a Wiener process with finite variance, which is
the most ``aggressive'' stochastic excitation still leaving problem
\eqref{eq:burgers} well-posed in $H^1_p$ (precise definition is
deferred to Section \ref{sec:noise}).  We add that the stochastic
Burgers equation {is related to} the Kardar-Parisi-Zhang equation,
which has received some attention in the literature
{\cite{h14a,mkv16}, except that the latter is typically studied in the
  presence of noise which is white both in space and in time}.

We now briefly summarize important results from the literature
relevant to the stochastic Burgers equation. The existence and
uniqueness of solutions has been proven in \cite{BCJL94,GN99} for the
problem posed on the real line and in \cite{DPDT94,g98} for a bounded
domain with Dirichlet boundary conditions. In all cases, solutions can
be regarded as continuous $L^p$-valued random processes. For the
bounded domain (the case which we are interested in), convergence of
numerical schemes has been established in \cite{AG06} for the
finite-difference approaches and in \cite{BJ13} for Galerkin
approximations.  However, in both cases only Dirichlet boundary
conditions were considered. The case with the periodic boundary
conditions has been recently considered in \cite{HM15} for a larger
class of Burgers-type equations and an abstract numerical scheme.

{There exists a large body of literature devoted to investigations of
  stochastically forced Burgers equation used as a model for
  three-dimensional (3D) turbulence. Below we mention a few landmark
  studies and refer the reader to the survey paper \cite{bk07} for
  additional details and references. The majority of these
  investigations aimed to characterize the solutions obtained in
  statistical equilibrium, attained by averaging over sufficiently
  long times, in terms of properties of the stochastic forcing. Given
  the motivation to obtain insights about actual turbulent flows, the
  main quantities of interest in these studies were the scaling of the
  energy spectrum, evidence for intermittency in the anomalous scaling
  of the structure functions and the statistics of $\partial_x u$,
  such as the tails (exponential vs.~algebraic) of its probability
  density function \cite{cy95a,cy95b,ztg97}. Remarkably, some of these
  results were also established with mathematical rigour \cite{b14}.
  The aforementioned quantities were also studied in flows evolving
  from stochastic initial data \cite{gk93}. In this context we mention
  the investigations \cite{s92,saf92} which focused on the statistics
  of shock waves in the limit of vanishing viscosity $\nu$.  As
  regards technical developments, a number of interesting results were
  obtained using optimization-based instanton formulations
  {\cite{mkv16,bfkl97,ggs15}}. While most of earlier investigations of
  stochastic problems in hydrodynamics were concerned with the
  properties of the statistically steady state obtained in the
  long-time limit \cite{ks12}, the focus of the present investigation
  is fundamentally different, as here we explore extreme forms of the
  transient behavior under stochastic excitation.  In other words,
  instead of studying the behavior in a time-averaged sense, we seek
  to understand how the worst-case scenarios are affected by
  stochastic forcing. The idea that stochastic excitation could act to
  re-establish global well-posedness in a system exhibiting a
  finite-time blow-up in the deterministic setting has been considered
  for some time, although more progress has been made on the related
  problem of restoring uniqueness \cite{f15}. The rationale for why
  noise might prevent the formation of singularities is that in some
  situations blow-up may require a simultaneous occurrence of certain
  conditions (phenomena) and this coincidence may be disrupted by
  stochastic excitations. There are in fact some model problems where
  such mechanism of regularization by noise has been proved to exist,
  including certain transport equations \cite{f15} and some versions
  of the Schr\"odinger equation \cite{ddm02a}. While there are a few
  related results available for the 3D Navier-Stokes and Euler
  equations \cite{f15b}, here we mention the studies \cite{ar09,ar10}
  where it was shown that singularity formation (gradient blow-up) in
  the inviscid Burgers equation can be prevented by a certain
  stochastic excitation of the associated Lagrangian particle
  trajectories.  }

However, there are also cases in which noise may amplify
  formation of singularities. For example, the paper \cite{ekms00}
  deals with the stochastic inviscid Burgers equation in which the
  stochastic forcing is periodic in the spatial coordinate and
  represented by white noise in time. The authors show that
  introducing noise increases the number of shocks present in the
  stochastic solution as compared to the deterministic case. In
  particular, this means that solutions are discontinuous (at almost
  all times $t$) and belong in a space of locally integrable
  functions. We will return to these results at the end of the paper.

For deterministic systems which exhibit blow-up, singularity formation
is typically signalled by unbounded growth of certain Sobolev norms
\cite{ks15a}. This growth can often be estimated using bounds obtained
with methods of functional analysis {and} even for problems which are
globally well-posed, such as the viscous Burgers equation
{\cite{kl04}}, it is important to understand how much {these Sobolev
  norms} can grow depending on the ``size'' of the initial data as
this can provide valuable insights concerning the sharpness of the
corresponding estimates. These issues are at the heart of the recently
undertaken research program aiming to probe the sharpness of
fundamental estimates on the growth of quadratic quantities in
hydrodynamic models \cite{ap11a,ap13a,ap13b}.  These estimates are of
two types, namely, concerning the instantaneous growth (i.e., the rate
of change at a fixed instant of time) and growth over finite time
windows. Important progress has also been made on some related
questions in the context of the 3D Navier-Stokes problem
\cite{ld08,ap16} which is in fact what has motivated this research
program.

\begin{figure}
  \centering
\subfigure[]{\includegraphics[width=.45\textwidth]{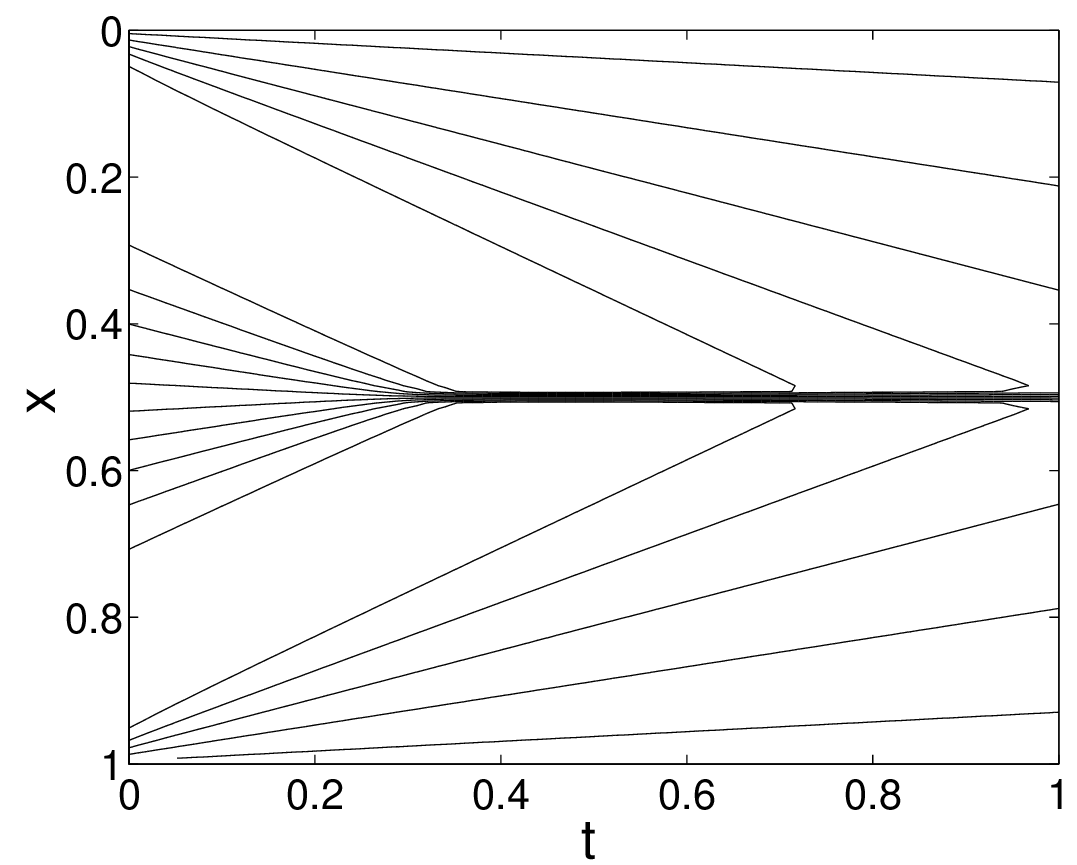}} \quad
\subfigure[]{\includegraphics[width=.45\textwidth]{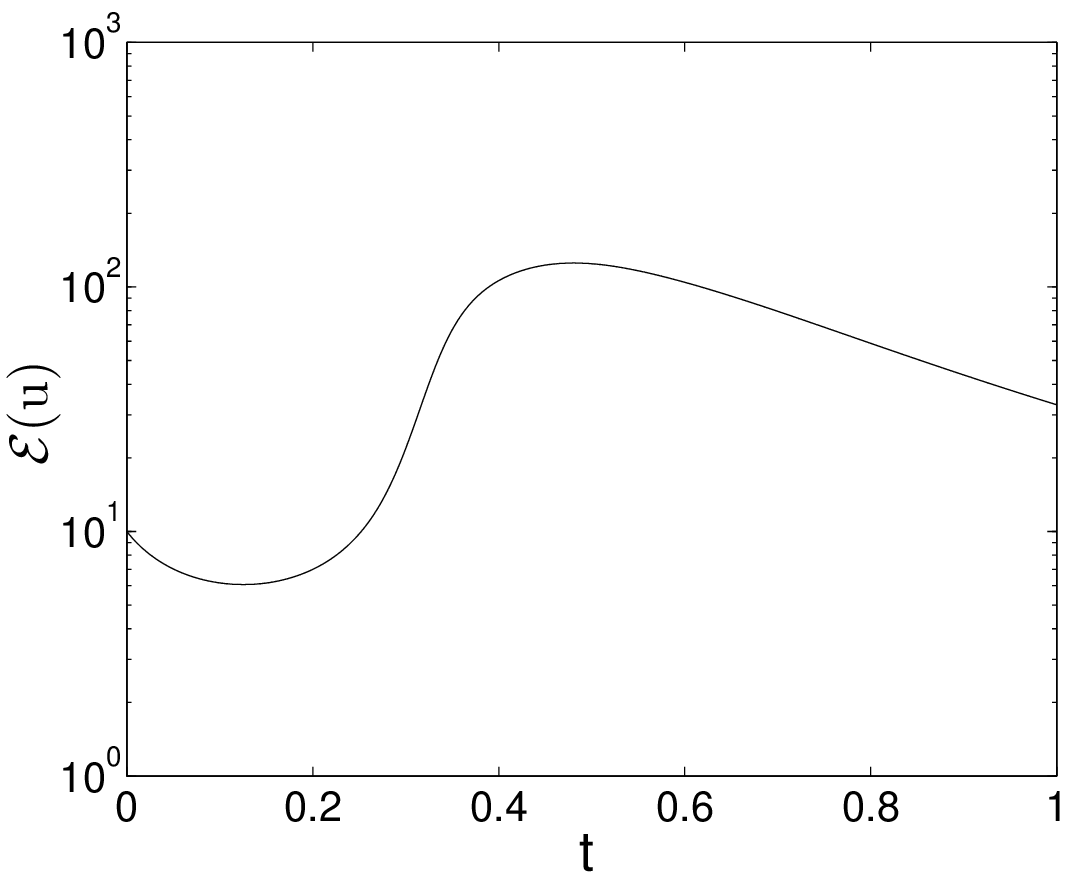}}
  \caption{(a) Space-time evolution of the solution $u(t,x)$ and (b)
    history of the enstrophy ${\E(u(t))}$ in a solution of the
    deterministic Burgers equation with an extreme initial condition
    $\tg0$. In figure (a) the level sets of $u(t,x)$ are plotted
      with increments of 0.1.}
    \label{fig:det}
\end{figure}
{For Burgers equation the key quantity of interest is} the $H^1$
seminorm of the solution referred to as {\em enstrophy}
\begin{equation}
  {\E(u(t))} := \frac{1}{2} \int_0^1 |\partial_x u(t,x)|^2 \, dx.
\label{eq:E}
\end{equation}
In the deterministic setting ($\zeta \equiv 0$ in
\eqref{eq:burgersA}), where Burgers equation is known to be globally
well-posed \cite{kl04}, its solutions generically exhibit a steepening
of the gradients (driven by the nonlinearity) followed by their
viscous dissipation when the linear dissipative term starts to
dominate. This behavior is manifested by an initial growth of
enstrophy ${\E(u(t))}$, which peaks when the solution $u(t,\cdot)$
builds up the steepest front, followed by its eventual decay to zero.
As a point of reference, we illustrate this generic behavior in Figure
\ref{fig:det} in which the results were obtained by solving system
\eqref{eq:burgers} with $\zeta \equiv 0$, {$T=1$} and an ``extreme''
initial condition $\tg0$ designed to produce a {\em maximum} enstrophy
growth over {$[0,1]$} for a given $\E_0 := {\E(\tg0)}$ \cite{ap11a}
(the numerical approach used to obtain the results in Figure
\ref{fig:det} and the construction of the extreme initial data $\tg0$
will be described {in Sections \ref{sec:numer} and \ref{sec:results},
  respectively}).  Although the evolution shown in Figure
\ref{fig:det} corresponds to a special choice of the initial data, it
is qualitatively similar to the generic case. {While the best estimate
  for the finite-time growth of enstrophy predicts $\max_{t \ge 0}
  \E(t) \le C\, \E_0^3$ for some $C>0$, where $\E_0 := \E(0) =
  {\frac{1}{2}\int_0^1 |\partial_x g(x)|^2 \, dx}$,
  computational evidence was presented in \cite{ap11a} that
  this estimate may not in fact be sharp and the largest
  possible growth of enstrophy actually scales as $\max_{t \ge 0}
  \E(t) \sim \E_0^{3/2}$. Given the relation between the growth of
  their Sobolev norms and the extreme (in particular, singular)
  behavior of solutions, it is important to understand whether this
  growth of enstrophy may be affected by stochastic excitation. The
  main goal of the present study is {therefore} to address this
  question in the context of the 1D Burgers equation.  In order to do
  so, we will have to use a more ``aggressive'' form of stochastic
  excitation than was used in earlier investigations of the stochastic
  Burgers problem where the forcing acted mostly on large scales.
  While this question is clearly of mathematical nature, in the
    absence of any theoretical estimates available for the effect of
    noise, either instantaneously or in finite time, on the growth of
    Sobolev norms of solutions to evolutionary stochastic PDEs, we
    will address it here through a series of carefully designed and
    executed computational experiments. The intention is that these
    results may motivate and guide further mathematical analysis of
    this problem. We add that, with the exception of the study
  \cite{ddm02a}, which concerned the stochastic Schr\"odinger
  equation, to the best of our knowledge there have been no
  computational studies of such problems.

\subsection{Summary of the Main Results}
\label{sec:summary}

The main question we address here is how the growth of the enstrophy
described by stochastic system \eqref{eq:burgers}, both in terms of
individual trajectories and statistical properties, depends on
the properties of the noise term in equation \eqref{eq:burgersA}, in
particular, whether this growth is enhanced or weakened in comparison
to the growth observed in the deterministic system \cite{ap11a},
cf.~Figure \ref{fig:det}. We have made the following observations:
\begin{itemize}
\item individual samples of the stochastic solution tend to
  exhibit a larger growth of enstrophy than the deterministic
  solution,

\item when the noise magnitude is sufficiently large relative to the
  initial enstrophy $\Ecal_0$, the dynamics of individual
  sample solutions is entirely dominated by noise and exhibits little
  effect of the initial data,

\item when the noise magnitude is small relative to the initial
  enstrophy, individual solution samples can be regarded as
  ``perturbations'' of the deterministic evolution with enstrophy
  growth dependent on $\Ecal_0$,

\item in statistical terms, the enstrophy growth
  $\max_{t\in[0,T]}\E({t})$ in the deterministic case provides
\begin{itemize}
\item an upper bound for the growth of the enstrophy of the expected
  value $\max_{t\in[0,T]}\E(\ebb[u(t)])$, and
\item a lower bound for the growth of the expected value of the
  enstrophy $\max_{t\in[0,T]}\ebb[\E(u(t))]$,
\end{itemize}

\item when the noise magnitude increases proportionally to the initial
  enstrophy $\Ecal_0$, the same growth of the expected
    value of the enstrophy is observed as in the deterministic
  case; this leads us to conclude that inclusion of
    stochastic forcing does not trigger any new mechanisms of
    enstrophy amplification.
\end{itemize}

The remainder of this paper is divided as follows: in the next section
we describe our model of noise and discuss some properties of the
stochastic solutions; the numerical approach is introduced {briefly}
in Section \ref{sec:numer}, whereas the computational results are
presented and discussed in {Section \ref{sec:results}}; conclusions
are presented in Section \ref{sec:final}.

\section{Structure of the Stochastic Forcing and Properties of the Solution}
\label{sec:noise}

As is customary in the standard theory of stochastic partial
differential equations (SPDEs), we write the stochastic Burgers
equation \eqref{eq:burgersA} in the differential form \cite{lps14}
\begin{equation}\label{eq:burdif}
du=\left(\nu\partial^2_xu{-}\frac{1}{2}\partial_xu^2\right)\, dt+\sigma\, dW,
\end{equation}
where $\zeta=\sigma\frac{dW}{dt}$ in which $\sigma>0$ is a constant
and $W(t)$ is a cylindrical Wiener process.  One can consider
  different notions of solution of system \eqref{eq:burgers}. Due to
  the lack of smoothness of the noise term, we do not expect to obtain
  solutions defined in the classical sense (i.e., solutions
  continuously differentiable with respect to the independent
  variables). One can, however, define the notion of a \emph{mild
    solution} as in \cite{DPDT94}
\begin{equation}
\label{eq:burint}
u(t)=e^{tA}g{-}\frac{1}{2}\int_0^te^{(t-s)A}\partial_x u^2\,ds+\sigma\int_0^t e^{(t-s)A}\,dW(s),
\end{equation}
where $A := \nu\partial_x^2$ and the action of $e^{tA}$ on $L^2$
functions is determined by the identity
\begin{equation*}
e^{tA} \, e^{2\pi i kx} = e^{-4\pi^2\nu tk^2} \, e^{2\pi i kx}, \qquad k\in\zbb, \quad x\in[0,1].
\end{equation*}
We remark that other notions of solution also exist, for example, the
notion of a \emph{weak solution} as defined in {\cite{AG06}.}

As regards the structure of the stochastic forcing,
$\{W(t)\}_{t\geq 0}$ is formally given by
\begin{equation}
W(t)=\sum_{j\in\nbb}\gamma_j\beta_j(t)\chi_j,
\label{eq:W}
\end{equation}
where $\{\beta_j(t)\}_{j\in\nbb}$ are i.i.d standard Brownian motions,
{$\{\chi_j\}_{j\in\nbb}$ form a trigonometric orthonormal basis,
  i.e.,}
\begin{equation}
{
\chi_0=1, \quad \chi_{2j}=\sqrt{2}\cos(2\pi jx), \quad \chi_{2j-1}=\sqrt{2}\sin(2\pi jx), \qquad j>0
}
\label{eq:chi}
\end{equation}
and $\{\gamma_j\}_{j\in\nbb}$ are scaling coefficients. When $\forall
j$ $\gamma_j=1$, $W$ is an infinite-variance Wiener process and
$\zeta$ is Gaussian noise white in both time and space, which is
commonly used in investigations of SPDEs.  However, this choice is not
suitable for the present study, since we are interested here in the
effects of stochastic excitation on the enstrophy, cf.~\eqref{eq:E},
and, as is demonstrated below, this quantity is {in fact} not
  defined for the Gaussian noise white in space.

Suppose $u$ is a mild solution satisfying equation
  \eqref{eq:burint} with an infinite-variance noise $W$.  {For
    convenience, let $\phi_k:=e^{2\pi i kx}$, $k\in\zbb$, denote
    {elements of the orthonormal Fourier basis}. We now study each of
    the terms appearing on the right-hand side of \eqref{eq:burint}.}

\textbf{Analysis of first term}: under the assumption that $g\in
L_p^2$, so that $\sum_{k\in\zbb}|\hat{g}_k|^2<\infty$, we have that
$e^{tA}g\in H_p^1$ (actually $H_p^\ell$ for any $\ell\geq 0$),
since
\begin{equation*}
\|e^{tA}g\|_{H_p^1}^2=\sum_{k\in\zbb}\frac{1+4\pi^2k^2}{e^{4\nu\pi^2k^2t}}|\hat{g}_k|^2<\infty
\end{equation*}
which is true because the exponentials dominate all other factors.

\textbf{Analysis of second term}: under the assumption that $u\in
L^2(\Omega,C([0,T],L_p^4))$, so that $u^2\in
L^2(\Omega,C([0,T],L_p^2))$, we can write
\begin{equation*}
u^2=\sum_{k\in\zbb}\hat{y}_k\phi_k \quad 
\text{with} \quad \sum_{k\in\zbb}\|\hat{y}_k\|_{L^2(\Omega,C([0,T],\cbb))}^2
=\sum_{k\in\zbb}\ebb\left[\sup_{0\leq t\leq T}|\hat{y}_k|^2\right]<\infty,
\end{equation*}
so that
\begin{equation*}
\partial_x u^2= 2\pi i \sum_{k\in\zbb} k\hat{y}_k\phi_k
\end{equation*}
and
\begin{equation*}
\int_0^t\frac{1}{2}e^{(t-s)A}\partial_x u^2\,ds
=\sum_{k\in\zbb} \left[\int_0^t\frac{2\pi i}{2}ke^{-4\nu\pi^2k^2(t-s)}\hat{y}_k(s)\, ds\right] \phi_k;
\end{equation*}
now each of the coefficients in the sum above can be bounded as
\begin{align*}
\left|\int_0^t\frac{2\pi i}{2}ke^{-4\nu\pi^2k^2(t-s)}\hat{y}_k(s)\,ds\right|&\leq\int_0^t\pi ke^{-4\nu\pi^2k^2(t-s)}ds \sup_{0\leq t\leq T}|\hat{y}_k|\\
&=\frac{1}{\nu\pi k}\left(1-e^{-8\nu\pi^2k^2t}\right)\sup_{0\leq t\leq T}|\hat{y}_k|\leq\frac{1}{\nu\pi k}\sup_{0\leq t\leq T}|\hat{y}_k|,
\end{align*}
so that the second term in \eqref{eq:burint} is also
in $H_p^1$ with
\begin{align*}
\left\|\int_0^t\frac{1}{2}e^{(t-s)A}\partial_x u^2\,ds\right\|^2_{L^2(\Omega,H_p^1)}&\leq\sum_{k\in\zbb}(1+4\pi^2k^2)\left\|\frac{1}{\nu\pi k}\sup_{0\leq t\leq T}|\hat{y}_k|\right\|_{L^2(\Omega,\cbb)}^2\\
&=\sum_{k\in\zbb}\frac{1+4\pi^2k^2}{\nu^2\pi^2k^2}\|\hat{y}_k\|_{L^2(\Omega,C([0,T],\cbb))}^2<\infty
\end{align*}
which follows from the summability of $\|\hat{y}_k\|^2$.

\textbf{Analysis of third term}: writing it in terms of a
Fourier series
\begin{equation*}
\sigma\int_0^t e^{(t-s)A}\,dW(s)=\sum_{k\in\zbb}\hat{W}_k(t)\, \phi_k,
\end{equation*}
we obtain (for $k>0$ with the cases $k=0$ and $k<0$
handled similarly)
\begin{equation*}
\hat{W}_k(t)=
\sigma\int_0^t e^{-4\nu\pi^2k^2(t-s)}\left(\frac{\sqrt{2}}{2}d\beta_{2k}(s)-i\frac{\sqrt{2}}{2}d\beta_{2k}(s)\right)
\end{equation*}
which is a random variable with the second moment given by
\begin{equation*}
\|\hat{W}_k(t)\|_{L^2(\Omega,\cbb)}^2
=\sigma^2\int_0^t e^{-8\nu\pi^2k^2(t-s)}\,ds=\frac{\sigma^2}{8\nu\pi^2k^2}\left(1-e^{-8\nu\pi^2k^2t}\right);
\end{equation*}
from this we see that the third term is in $L^2$ but not
in $H_p^1$, as for any $t>0$ we have
\begin{equation*}
\begin{aligned}
\left\|\sigma\int_0^t e^{(t-s)A}\,dW(s)\right\|_{L^2(\Omega,H_p^1)}^2
& =\sum_{k\in\zbb}(1+4\pi^2k^2)\|\hat{W}_k\|_{L^2(\Omega,\cbb)}^2 \\
& =\frac{\sigma^2}{8\nu\pi^2}\sum_{k\in\zbb}\frac{1+4\pi^2k^2}{k^2}(1-e^{-8\nu\pi^2k^2t})=\infty.
\end{aligned}
\end{equation*}
We therefore conclude that while the first two terms on the right-hand
side of \eqref{eq:burint} are in $H_p^1$ (and hence also in $L^2$),
the third one is only in $L^2$ and not in $H_p^1$. Thus, for any $t >
0$, $u(t)$, being the left-hand side of \eqref{eq:burint}, is in $L^2$
but not in $H_p^1$, and consequently the enstrophy obtained with
Gaussian noise white in space is not well defined.

We shall thus focus on noise
representations with $\ell^2$-summable coefficients, such as
\begin{equation}
\label{eq:sclgam}
\gamma_0=1, \quad \gamma_{2k+1}=\gamma_{2k+2}=\frac{1}{k}, \qquad k>0,
\end{equation}
so that $W(t)$ has a {\em finite} variance, meaning that it is
square-integrable in $L^2$, i.e., $W(t)\in L^2(\Omega,L^2)$, with the norm
\begin{equation}
\|W(t)\|_{L^2(\Omega,L^2)}^2 
= \sum_{j\in\nbb}|\gamma_j|^2\|\beta_j(t)\|_{L^2(\Omega,\cbb)}^2\|\chi_j\|_{L^2}^2
= t\sum_{j\in\nbb}\gamma_j^2=\left(1+\frac{\pi^2}{3}\right)t.
\label{eq:Wnorm}
\end{equation}
{Such a finite-variance Wiener process ensures that the enstrophy
  is a well-defined quantity. The corresponding term $\zeta$ in
  equation \eqref{eq:burgersA} will be referred to as the
  \emph{Gaussian colored-in-space noise}. We add that a finite-variance Wiener
  process may also be constructed with scaling coefficients
  $\{\gamma_j\}_{j\in\nbb}$ decaying a bit less rapidly than indicated
  in \eqref{eq:sclgam}, namely as
  $\gamma_{2k+1}=\gamma_{2k+2}=1/k^{1/2+\epsilon}$ or
  $\gamma_{2k+1}=\gamma_{2k+2}=(\ln k)^{1+\epsilon}/k$, $k>0$, for
  some $\epsilon > 0$. We tested stochastic actuation with such
  structure computationally, but in terms of the quantities we are
  interested in there was no appreciable difference with respect to
  \eqref{eq:sclgam}. Therefore, hereafter we will focus on stochastic
  excitations defined by \eqref{eq:sclgam}.}

\section{Numerical Approach}
\label{sec:numer}

System \eqref{eq:burgers} will be discretized with respect to the
three independent variables, namely, the space variable $x$, time $t$
and the stochastic variable $\omega \in \Omega$.  Since our numerical
approach is fairly standard (similar techniques were employed in
\cite{cy95a,cy95b}), we describe it below only briefly.  The approach
is then validated in Section \ref{sec:validate}.

Discretization with respect to the space variable $x$ is performed
using a spectral approach based on truncated Fourier series.  {Since}
the nonlinear term {$(1/2)\partial_x u^2$} is represented as a
convolution sum in {the} Fourier space, {it can be evaluated more
  efficiently in the physical space} with a pseudospectral approach
based on the Fast Fourier Transforms (FFTs) combined with dealiasing
based on the ``3/2'' rule \cite{canuto:SpecMthd}. We let $K$ denote
the discretization parameter {equal} to the number of Fourier modes,
{so that} $K=\lfloor\frac{M}{3}\rfloor$, where $M$ is the number of
grid points {in the physical space. To maximize the performance of
  FFTs, $M$ will be taken} to be a large power of 2.

Discretization with respect to the time variable $t$ is performed
using a finite-difference approach based on a uniform grid in time.
We use a semi-implicit (first-order) Euler method in which the
dissipative term is treated implicitly, whereas the nonlinear and the
stochastic terms are treated explicitly. We let $N$ denote the
discretization parameter {representing} the number of time steps {in
  the interval $[0,T]$.}

Discretization {of} the stochastic forcing {$\zeta$} is performed
using a Monte Carlo approach to sample the distribution {of the
  stochastic variable $\omega \in \Omega$}. We compute realizations of
the stochastic solution for a sequence of noise samples {which,
  consistently with the spectral approach to discretization in space,
  are represented as random realizations of the coefficients
  $\{\beta_j(t)\}_{j\in\nbb}$ in Fourier expansion \eqref{eq:W}.}  The
expected values of the Fourier coefficients {of the solution} can then
be approximated using the \emph{average} estimator \cite[Section
4.4]{lps14}. We let $S$ denote the discretization parameter
{representing} the number of samples.

For $k=0,\ldots K$, $n=0,\ldots,N$ and $s=1,\ldots,S$, we let
$\hat{u}^{K,N,S}_{k,n,s}$ denote the $s$-th realization of the $k$-th
Fourier mode of $u =u(t,x;\omega)$ at time $t_n=\frac{n}{N}T$. We
recall that we wish to compute the enstrophy of the solution, defined
in \eqref{eq:E}.  In the stochastic setting, there are two distinct
quantities of interest: one can either consider the \emph{enstrophy of
  the expected value} of the {stochastic} solution, or the
\emph{expected value of the enstrophy} of the {stochastic} solution.
Estimates of both these quantities can be obtained {using} the
expressions
\begin{subequations}
\label{eq:Eest}
\begin{align}
\Ecal(\ebb[u(t_n)]) & \approx\sum_{k=1}^K4\pi^2k^2\left|\frac{1}{S}\sum_{s=1}^S\hat{u}^{K,N,S}_{k,n,s}\right|^2, \label{eq:Eest1} \\
\ebb[\Ecal(u(t_n))] & \approx\sum_{k=1}^K4\pi^2k^2\frac{1}{S}\sum_{s=1}^S\left|\hat{u}^{K,N,S}_{k,n,s}\right|^2. \label{eq:Eest2}
\end{align}
\end{subequations}
These two quantities (and also their estimates) are related via Jensen's inequality \cite{lps14}
\begin{equation}
\Ecal(\ebb[u(t)]) \; \le \; \ebb[\Ecal(u(t))]. 
\label{eq:jensen}
\end{equation}
The reason for also including $\Ecal(\ebb[u(t)])$ in our
  analysis is that quantities related to averaged (mean) fields are
  often employed in statistical theories of turbulent flows
  \cite{Pope2000book,davidson:turbulence}, e.g., in the context of the
  so-called Reynolds-Averaged Navier-Stokes equations.

Our choice of the Monte Carlo approach to noise sampling is motivated
by its well-understood convergence properties and straightforward
implementation. While more modern approaches, such as polynomial chaos
expansions, may in principle achieve faster convergence, they suffer
from much higher computational complexity (at least polynomial in the
number of random variables, which is $N(2K+1)$ for our
discretization).  Moreover, the nonlinear term will have a rather
complicated expression in the polynomial orthonormal basis, a
challenge which does not arise only in linear problems \cite[Chapter
9]{lps14}. We also remark that the low-order of the time-integration
scheme {in our approach} is justified by the need to simultaneously
account for stochastic excitation which is not a smooth function of
time.

{\subsection{Validation}
\label{sec:validate}

Since a rigorous convergence proof of the numerical approach presented
above would be outside the scope of the present study, we limit
ourselves to showing computational evidence that this approach is
indeed convergent. Given {that there are} three numerical parameters,
$M$, $N$ and $S$, this is achieved by studying solutions to problem
\eqref{eq:burint} as each of the three parameters is refined with the
other two held fixed. In each case we monitor the difference between
approximations of the quantities \eqref{eq:Eest1} and \eqref{eq:Eest2}
and their values corresponding to the reference solution obtained with
the finest discretization: $K=341=\lfloor\frac{1024}{3}\rfloor$
dealiased complex Fourier modes (corresponding to $M = 1024$ grid
points in the physical space), $N=20,000$ time steps and $S = 1000$
Monte Carlo samples. These results, obtained using the initial data
$\tg0$ with $\E_0 = 10$ and $T = 1$, are presented in Figures
\ref{fig:nconv}a,b,c. In Figure \ref{fig:nconv}a, showing the effect
of the spatial discretization parameter $M$ with $N=20,000$ and
$S=1000$ fixed, we see that the rate of decrease of the error
increases with $M$, which is an indication of spectral convergence. In
Figure \ref{fig:nconv}b, showing the effect of the temporal
discretization parameter $N$ with $M = 1024$ and $S=1000$ fixed, we
observe linear convergence of the errors for both quantities of
interest. Finally, in Figure \ref{fig:nconv}c, showing the effect of
the stochastic sampling parameter $S$ with $M = 1024$ and $N=20,000$
fixed, we see that the rate of convergence of the error is about
$1/2$. We thus conclude that the proposed numerical approach is
convergent, with the expected rates of convergence \cite{lps14}, as
each of the three numerical parameters is refined. The numerical
parameters characterizing the reference solution described above
represent a reasonable trade-off between accuracy and computational
cost, and were used to obtain the results presented in Section
\ref{sec:results}.  To simplify the notation, we will use the symbol
$u = u(t,x;\omega)$ to represent the solution obtained numerically
with {these} parameter values.}

\begin{figure}
 \centering
\subfigure[]{\includegraphics[width=.45\textwidth]{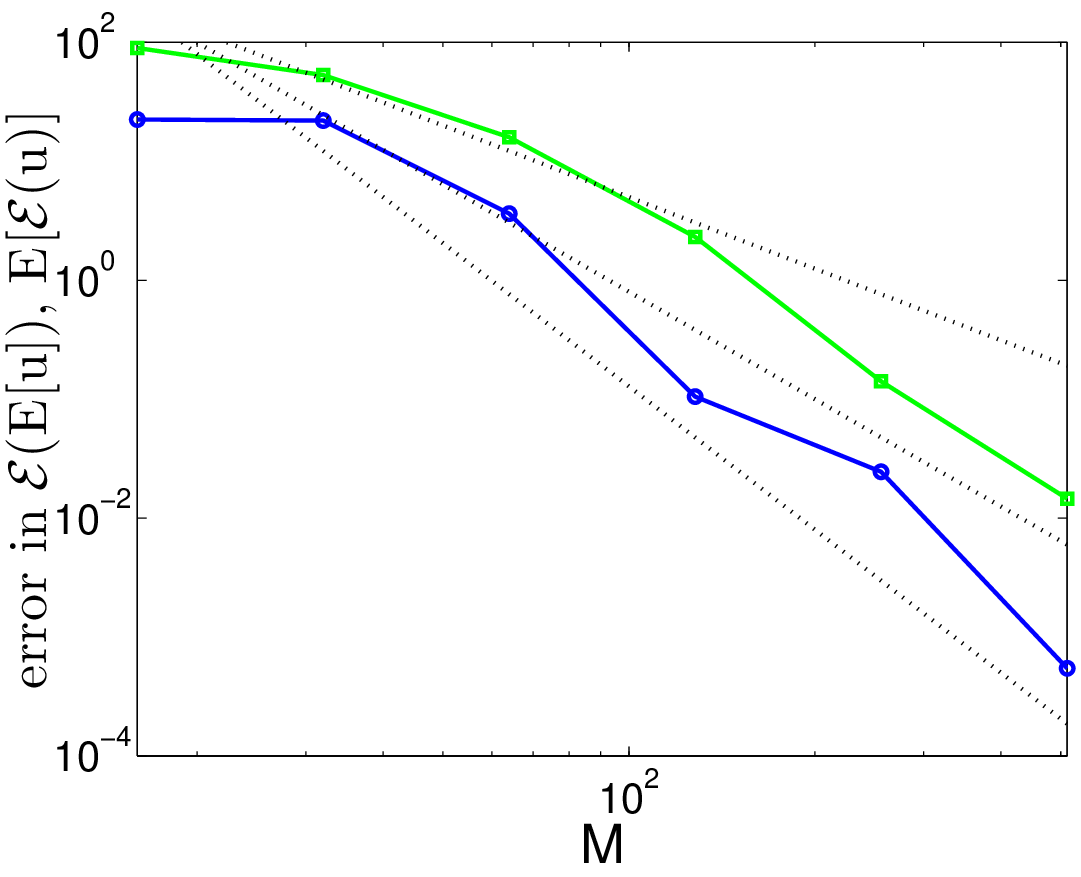}}\qquad
\subfigure[]{\includegraphics[width=.45\textwidth]{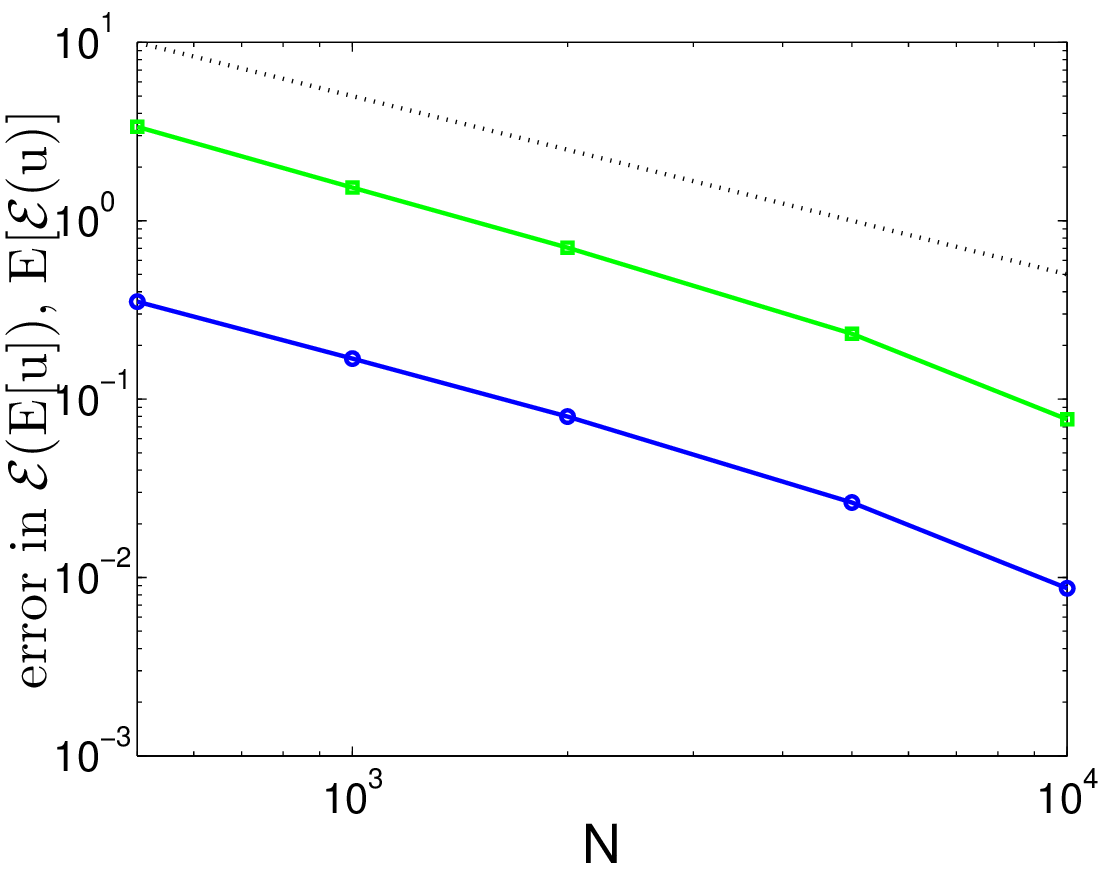}}
\subfigure[]{\includegraphics[width=.45\textwidth]{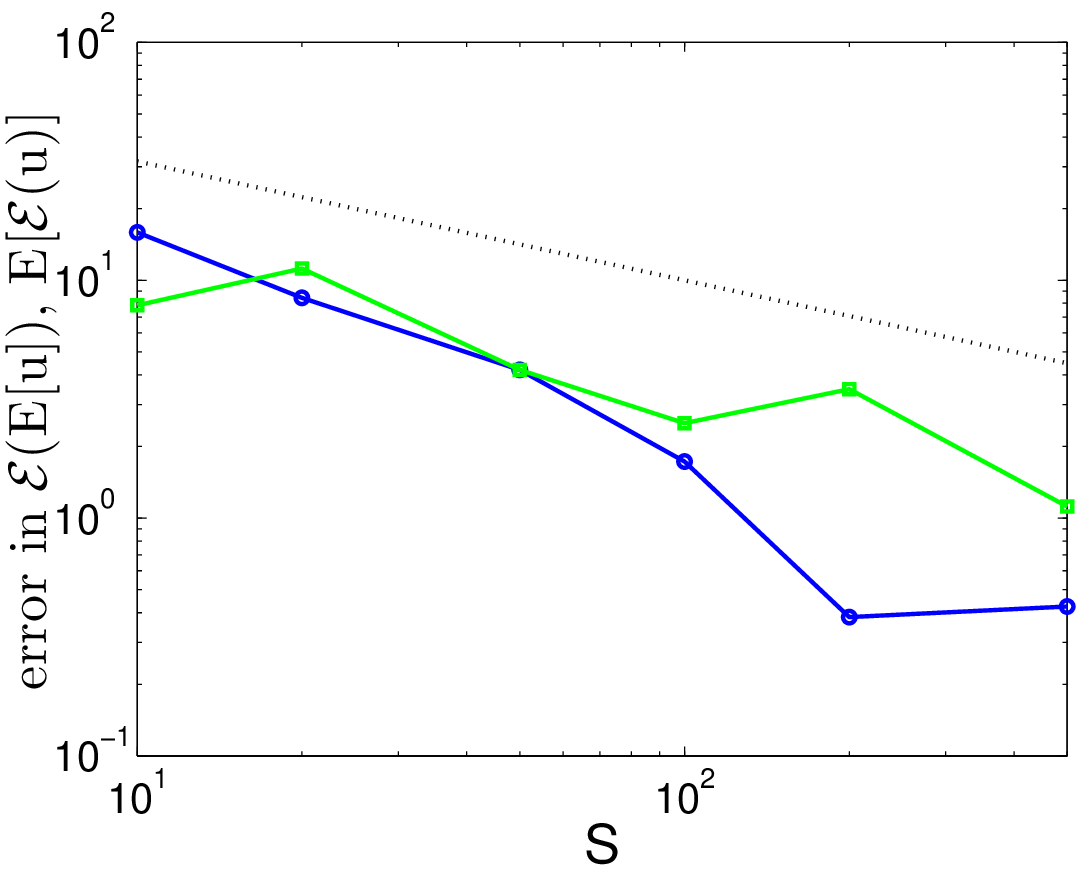}}
 \caption{{Errors in the numerical approximations of
     $\Ecal(\ebb[u(T)])$ (blue lines and circles) {and}
     $\ebb[\Ecal(u(T))]$ (green lines and squares) as functions of (a)
     the spatial discretization parameter $M$ with $N=20,000$ and
     $S=1000$ fixed, (b) the temporal discretization parameter $N$
     with $M = 1024$ and $S=1000$ fixed and (c) the sampling
     discretization parameter $S$ with $M = 1024$ and $N=20,000$
     fixed.  The initial data used was $\tg0$ with $\Ecal_0 = 10$ and
     $T=1$, and the errors are evaluated with respect to the reference
     solutions computed with $M=1024$, $N=20,000$ and $S=1000$. The
     dashed black lines correspond to the power laws (a) $CM^{-2}$,
     $CM^{-3}$, and $CM^{-4}$ , (b) $CN^{-1}$, and (c) $CS^{-1/2}$
     with suitably adjusted constants $C$.}}
 \label{fig:nconv}
\end{figure}

\section{Computational Results}
\label{sec:results}

In this section we use the numerical approach introduced above to
study the effect of the stochastic excitation with the structure
described in Section \ref{sec:noise} on the enstrophy growth in the
solutions of Burgers equation. More specifically, we will address the
question formulated in Introduction, namely, whether or not the
presence of noise can {change} the maximum growth of enstrophy
{observed} in the deterministic setting in \cite{ap11a}.  We will
do so by studying how the growth of the two quantities,
$\Ecal(\ebb[{u}])$ and $\ebb[\Ecal({u})]$ introduced in
Section \ref{sec:numer}, is affected by the stochastic excitation as a
function of the initial enstrophy $\Ecal_0 = \frac{1}{2} \int_0^1
|\partial_x g(x)|^2 \, dx$.  {Given time intervals of different
  length $T$, we} will solve system \eqref{eq:burgers} subject to {\em
  optimal} initial condition $\tg0$ which is designed to produce the
largest possible growth of enstrophy at time $T$ for all initial data
in $H^1_p$ with enstrophy $\Ecal_0$.  The procedure for obtaining such
optimal initial data is discussed in \cite{ap11a} and the optimal
initial conditions corresponding to $\Ecal_0 = 10$ and different time
windows $T$ are shown in Figure \ref{fig:g}. We see in this figure
that, as $T$ increases, the form of the optimal initial data changes
from a ``shock wave'' to a ``{rarefaction} wave''. {We remark
  that the optimal initial data $\tg0$ was obtained in the
  deterministic setting and, as such, might not produce optimal
  enstrophy growth in the presence of stochastic forcing. To probe
  such possibility, we also conducted tests with other initial
  conditions in the form $g(x) = A \sin(2\pi k x)$, where
  $k=1,2,\dots$ and $A \in \rbb$ was chosen to satisfy the condition
  $\E(g) = \E_0$. We note that for different values of $k$ such
  initial conditions represent mutually orthogonal ``directions'' in
  the space $H_p^1(0,1)$.  However, in all such cases the observed
  growth of $\Ecal(\ebb[{u}])$ and $\ebb[\Ecal({u})]$ was
  always inferior to the growth obtained with the initial data $\tg0$,
  hence these results are not reported here.}
\begin{figure}
  \centering
    \includegraphics[width=0.5\textwidth]{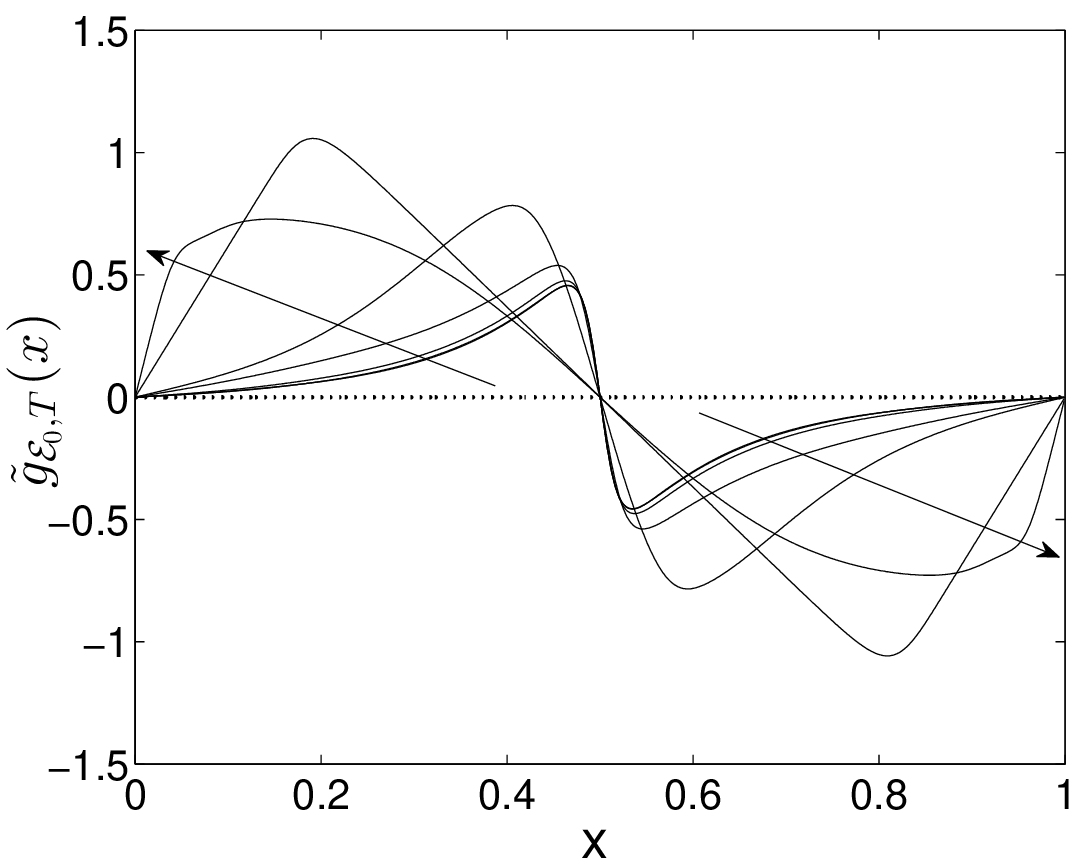}
    \caption{Optimal initial conditions $\tg0(x)$ for $\Ecal_0=10$ and
      $T$ ranging from $10^{-3}$ to $1$ \cite{ap11a} (arrows indicate
      the directions of increase of $T$).}
\label{fig:g}
\end{figure}

In the subsections below we first recall some
properties of the extreme enstrophy growth in the deterministic
setting and then discuss the effect of the noise on the enstrophy
growth over time and globally as a function of $\Ecal_0$.

\subsection{Deterministic Case Revisited}
\label{sec:deterministic}

The deterministic case will serve as a reference and here we summarize
some key facts about the corresponding maximum enstrophy growth. The
reader is referred to studies {\cite{ap11a,ld08,p12,p12b}} for
additional details. As illustrated in Figure \ref{fig:det}, a typical
behavior of the solutions to Burgers equation involves a steepening of
the initial gradients, which is manifested as a growth of enstrophy,
followed by their dissipation when the enstrophy eventually decreases.
The key question is how the enstrophy at some fixed time $\Ecal(T)$,
or the maximum enstrophy $\max_{t \in [0,T]} \Ecal(t)$, depend on the
initial enstrophy $\Ecal_0$. While the sharpest available analytical
estimate predicts $\max_{t \ge 0} \Ecal(t) \le C\, \Ecal_0^3$ for
large $\Ecal_0$, it was found in \cite{ap11a} that under the most
extreme circumstances the actual system evolution does not saturate
this upper bound producing instead $\max_{t \in [0,T]} \Ecal(t) \sim
\Ecal_0^{3/2}$. These results are illustrated in Figure
\ref{fig:Edet}a,b, where we can also see that for very short evolution
times growth only linear in $\Ecal_0$ is observed (this is because for
small $\Ecal_0$ the solutions do not have enough time to produce sharp
gradients). Since for increasing $\Ecal_0$ the maximum growth of
enstrophy is achieved for different $T$, the power-law behavior is
obtained by taking a maximum of $\Ecal(T)$ or $\max_{t \in [0,T]}
\Ecal(t)$ with respect to $T$ (represented in Figures
\ref{fig:Edet}a,b as ``envelopes'' of the curves corresponding to
different values of $T$).
\begin{figure}
\subfigure[]{\includegraphics[width=.45\textwidth]{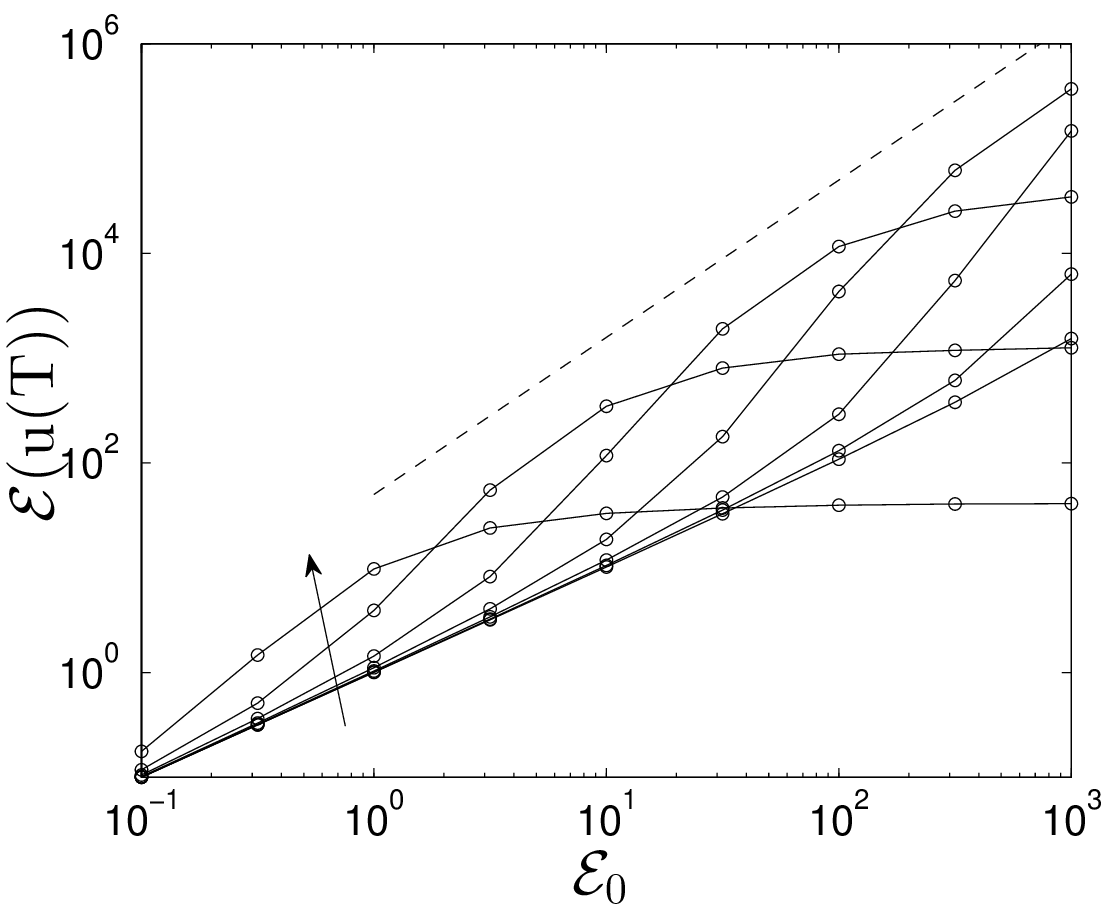}}\quad
\subfigure[]{\includegraphics[width=.45\textwidth]{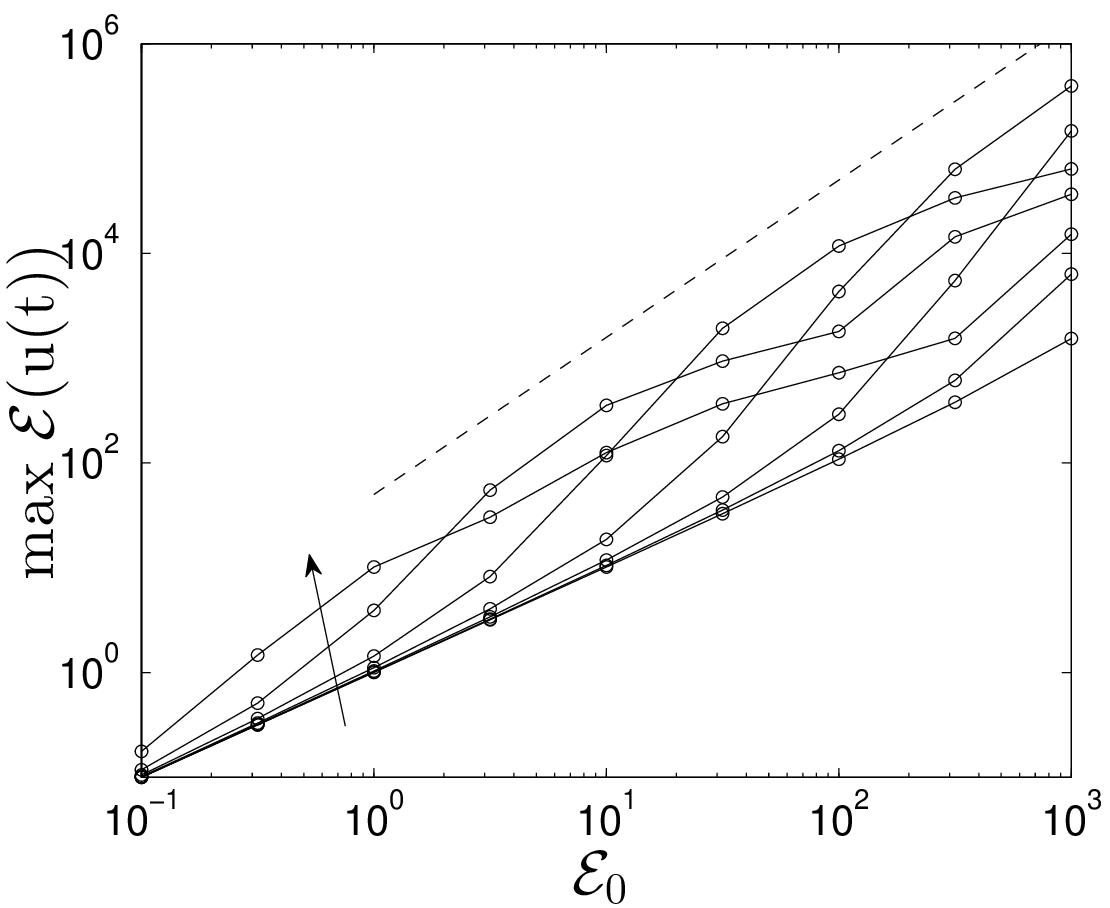}}
 \centering
  \caption{Dependence of (a) the enstrophy $\Ecal(T)$ at a final time
    $T$ and (b) the maximum enstrophy $\max_{t \in [0,T]} \Ecal(t)$ on
    the initial enstrophy $\Ecal_0$ for the optimal initial data
    $\tg0$ with {$T$ in the range from $10^{-3}$ to $1$}. Arrows indicate the
    direction of increasing $T$ and the {dashed} lines correspond to
    the power law $C \, \Ecal_0^{3/2}$.}
\label{fig:Edet}
\end{figure}

\subsection{Effect of Noise on Time Evolution}
\label{sec:effect_time}

We now analyze the effect of noise, both in terms of individual
trajectories and in the statistical sense, as a function of time
during the evolution starting from the optimal initial data $\tg0$
with {enstrophy $\Ecal_0 = 10$ and a fixed final time $T=1$}.
Stochastic solutions corresponding to ``small'' noise magnitude
$\sigma^2 = 10^{-2}$ and ``large'' noise magnitude {$\sigma^2 = 1$}
are illustrated in Figures \ref{fig:smpsml} and \ref{fig:smpbig},
respectively. The individual stochastic trajectories are shown as
functions of space and time in Figures \ref{fig:smpsml}a and
\ref{fig:smpbig}a. We see that in the small-noise case the effect of
the stochastic excitation is to gradually change the position of the
``shock wave'' (cf.~Figures \ref{fig:det}a and \ref{fig:smpsml}a).  In
the large-noise case the steep gradient region from the initial data
is gone and is replaced with spontaneously appearing and interacting
shocks which move in a largely structureless field (Figure
\ref{fig:smpbig}a). The corresponding evolutions of the enstrophy of
some sample stochastic solutions $\Ecal({u(t;\omega_s)})$,
$s=1,2$, the expected value of the enstrophy $\ebb[\Ecal({
  u(t)})]$ and the enstrophy of the expected value of the solution
$\Ecal(\ebb[{ u(t)}])$ are shown in Figures \ref{fig:smpsml}b and
\ref{fig:smpbig}b for the two noise levels where they are also
compared to the enstrophy evolution $\Ecal(t)$ in the deterministic
case. We see that the enstrophy of the sample stochastic solutions
tends to exceed the enstrophy of the deterministic solution for most,
albeit not all, times. As regards the relation of the the expected
value of the enstrophy $\ebb[\Ecal({ u(t)})]$ and the enstrophy
of the expected value of the solution $\Ecal(\ebb[{ u(t)}])$ to
the enstrophy $\Ecal(t)$ in the deterministic case, the following
relationship is observed
\begin{equation}
\Ecal(\ebb[{ u(t)}]) \ \leq \ \Ecal(t) \ \leq \ \ebb[\Ecal({ u(t)})], \quad t > 0
\label{eq:EEE}
\end{equation}
for both noise levels. While the relation between $\Ecal(\ebb[{
  u(t)}])$ and $\ebb[\Ecal({ u(t)})]$ is a consequence of
Jensen's inequality \eqref{eq:jensen}, the fact that these two
quantities in fact bracket the enstrophy of the deterministic solution
uniformly in time appears rather non-obvious.  This conclusion is
further elaborated in Figure \ref{fig:incsgm} where we show the time
evolution of the three quantities from \eqref{eq:EEE} for increasing
noise levels. We see that the difference between $\Ecal(\ebb[{
  u(t)}])$ and $\ebb[\Ecal({ u(t)})]$ increases with the noise
magnitude $\sigma^2$, such that at large noise levels the enstrophy of
the expected value of the solution exhibits no growth at all. The
fluctuations evident in $\Ecal(\ebb[{ u(t)}])$ corresponding to
the largest noise level are a numerical artefact resulting from an
insufficient number of Monte Carlo samples, {due to the fact that
  increased noise levels slow down the convergence of the Monte Carlo
  approach.}
\begin{figure}
  \centering
\subfigure[]{\includegraphics[width=.45\textwidth]{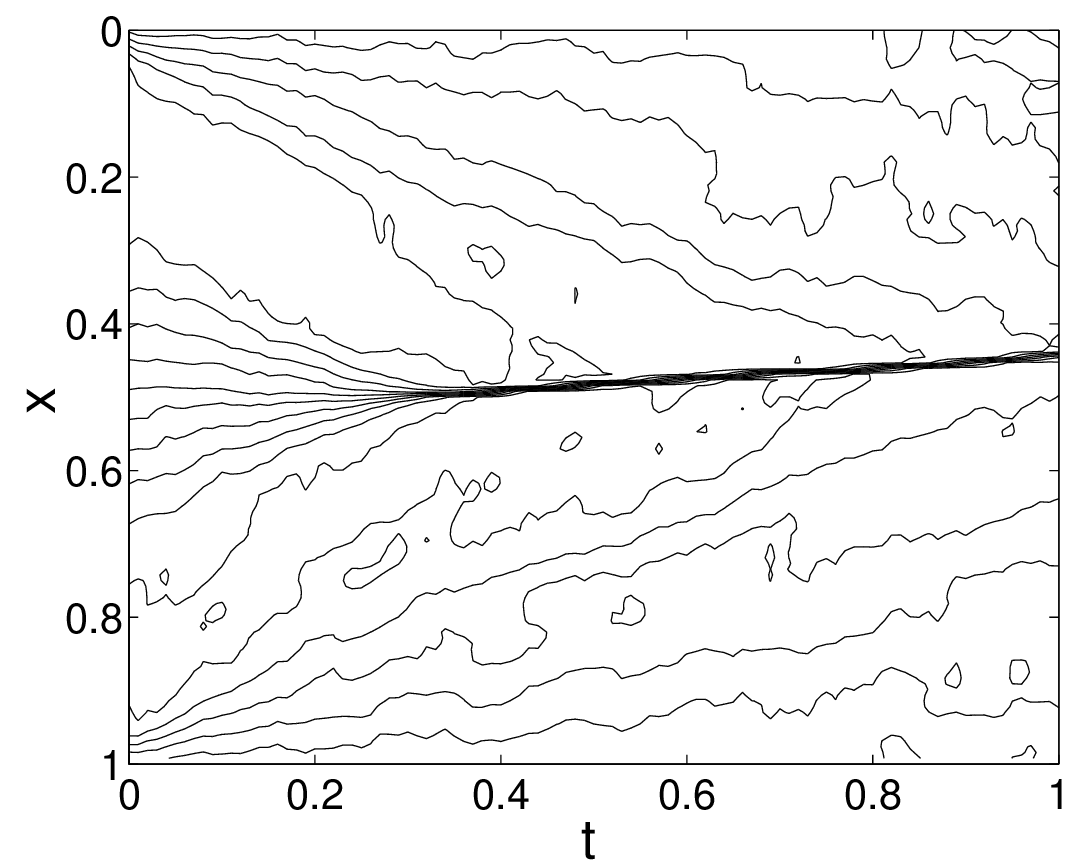}}\quad
\subfigure[]{\includegraphics[width=.45\textwidth]{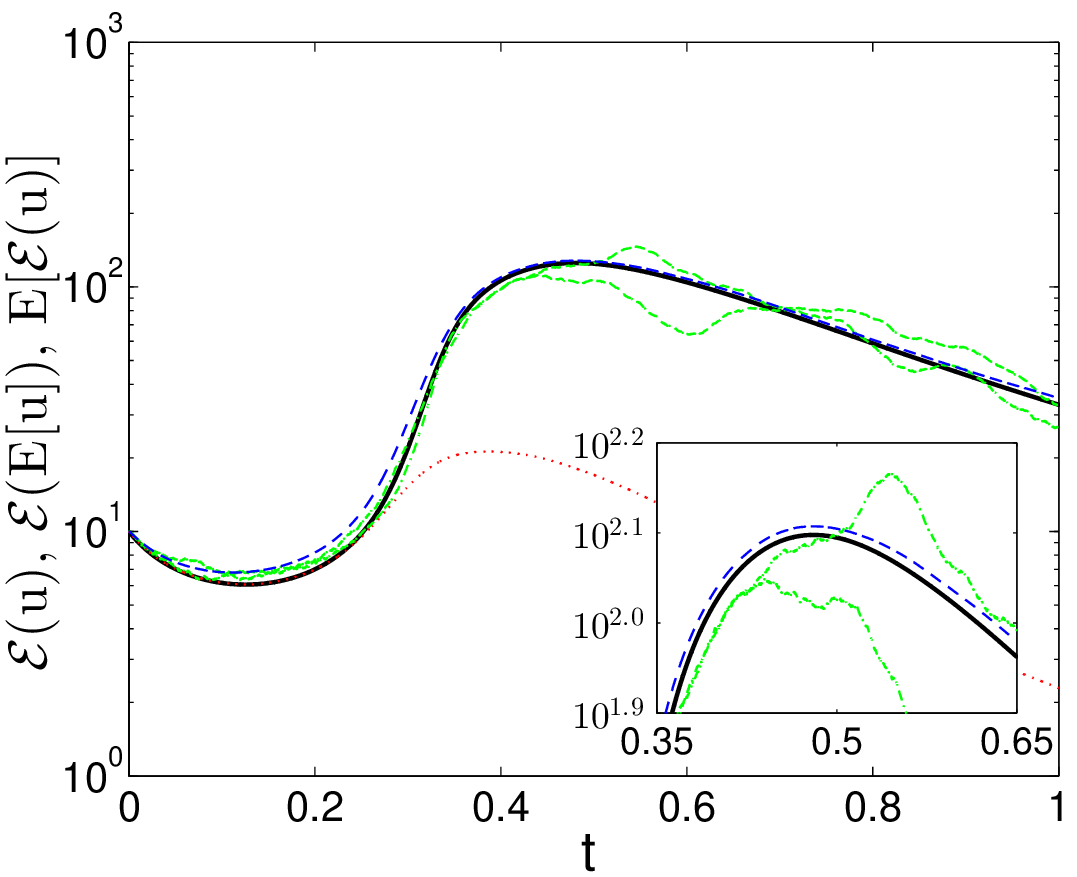}}
  \caption{[Small noise case: $\sigma^2 = 10^{-2}$] (a) Sample
    stochastic solution ${ u(t,x)}$ as a function of space and
    time (the level sets are plotted with the increments of {$0.1$}),
    (b) evolution of enstrophy of two sample stochastic solutions
    $\Ecal({ u(t;\omega_s)})$, $s=1,2$, (green dash-dotted lines), {the
      enstrophy of the deterministic solution $\Ecal(t)$ (black
      solid line),} the expected value of the enstrophy
    $\ebb[\Ecal({ u(t)})]$ (blue dashed line) and the enstrophy of
    the expected value of the solution $\Ecal(\ebb[{ u(t)}])$ (red
    dotted line). The initial data used was $\tg0$ with $\Ecal_0 = 10$
    and $T=1$. The inset in figure (b) shows details of the
      evolution during the subinterval $[0.35,0.65]$.}
  \label{fig:smpsml}
\bigskip\bigskip
  \centering
\subfigure[]{\includegraphics[width=.45\textwidth]{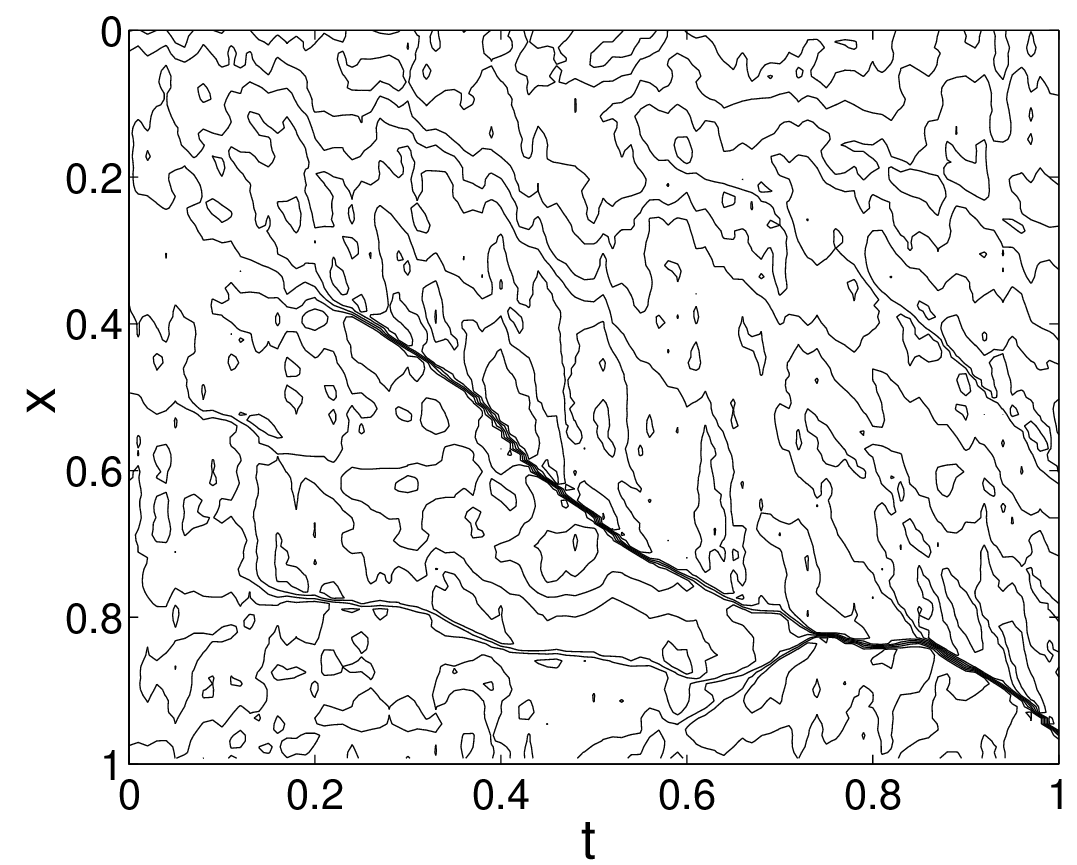}}\quad
\subfigure[]{\includegraphics[width=.45\textwidth]{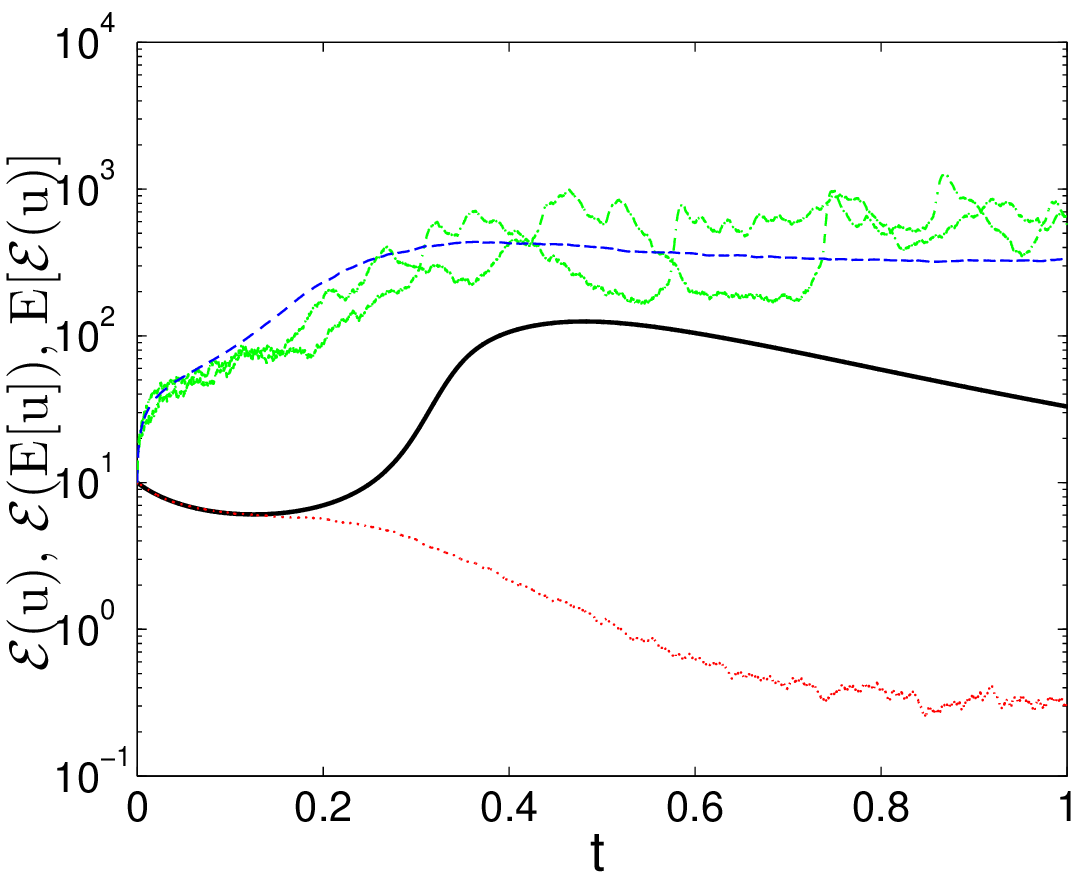}}
  \caption{[Large noise case: $\sigma^2=1$] (see previous figure for details).}
  \label{fig:smpbig}
\end{figure}

\begin{figure}
  \centering
    \includegraphics[width=0.45\textwidth]{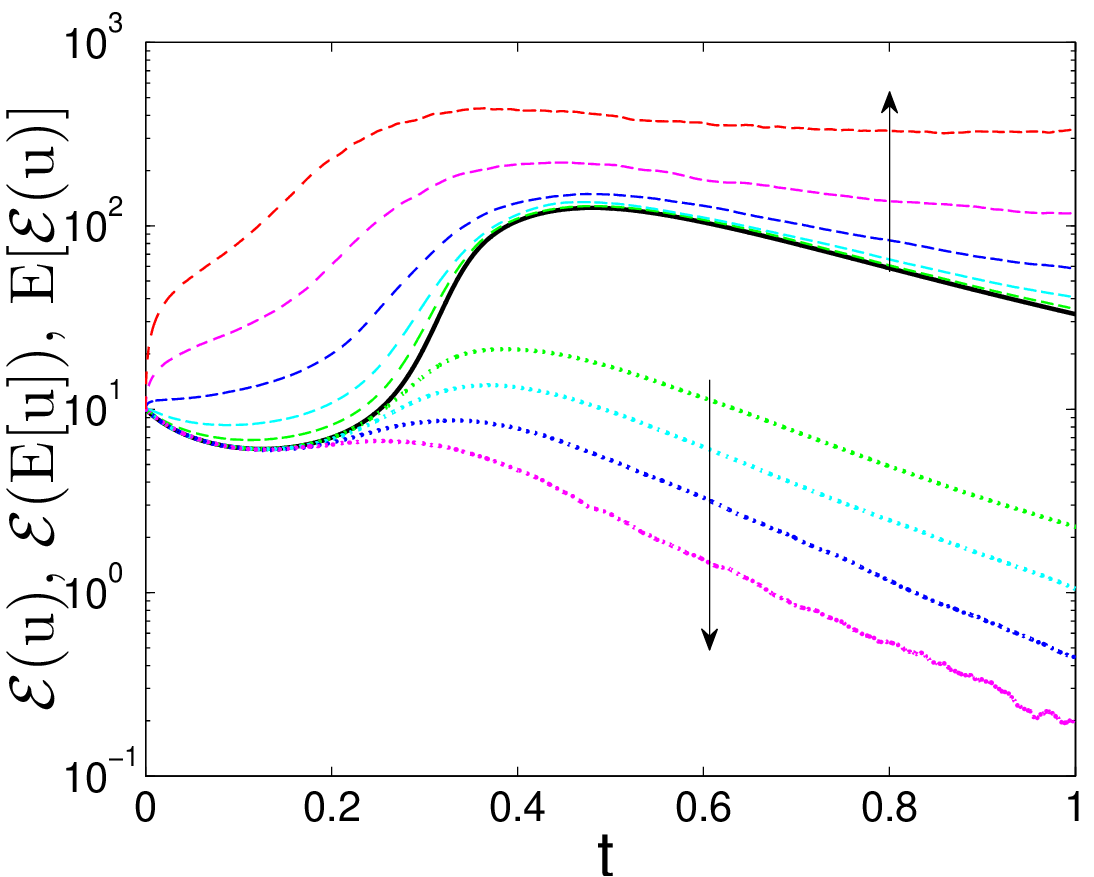}
  \caption{The expected value of the enstrophy
    $\ebb[\Ecal({ u(t)})]$ (dashed lines), the enstrophy of the
    expected value of the solution $\Ecal(\ebb[{ u(t)}])$ (dotted
    lines) and the enstrophy $\Ecal(t)$ of the deterministic solution
    (thick solid line) as functions of time for the initial condition
    $\tg0$ with $\Ecal_0 = 10$, $T=1$ and different noise levels
    $\sigma^2$ in the range from $10^{-2}$ to $1$ (the direction of
    increase of $\sigma^2$ is indicated by arrows).}
  \label{fig:incsgm}
\end{figure}

{The distributions of the maximum enstrophy values $\max_{t \ge
    0} \E(u({t;\omega}))$ corresponding to different stochastic
  realizations $\omega$ of the noise are shown for the cases with
  $\E_0 = 10, 10^3$ and $T = 1$ as probability distribution functions
  (PDFs) in Figures \ref{fig:pdf}(a,b). It is evident from these
  figures that the PDFs are non-Gaussian and, in particular, are
  asymmetric with heavy, possibly algebraic, tails characterizing
  values of $\max_{t \ge 0} \E({u(t,\omega)})$ larger than
  $\ebb[{\max_{t \ge 0} \Ecal(u(t))}]$. However, it is also clear that
  the deviation from the Gaussian behavior is significantly smaller in
  the larger enstrophy case (Figure \ref{fig:pdf}(b)) than in the
  lower enstrophy case (Figure \ref{fig:pdf}(a)). This deviation also
  tends to increase with the noise magnitude $\sigma^2$.}

\begin{figure}
  \centering
\subfigure[]{\includegraphics[width=.475\textwidth]{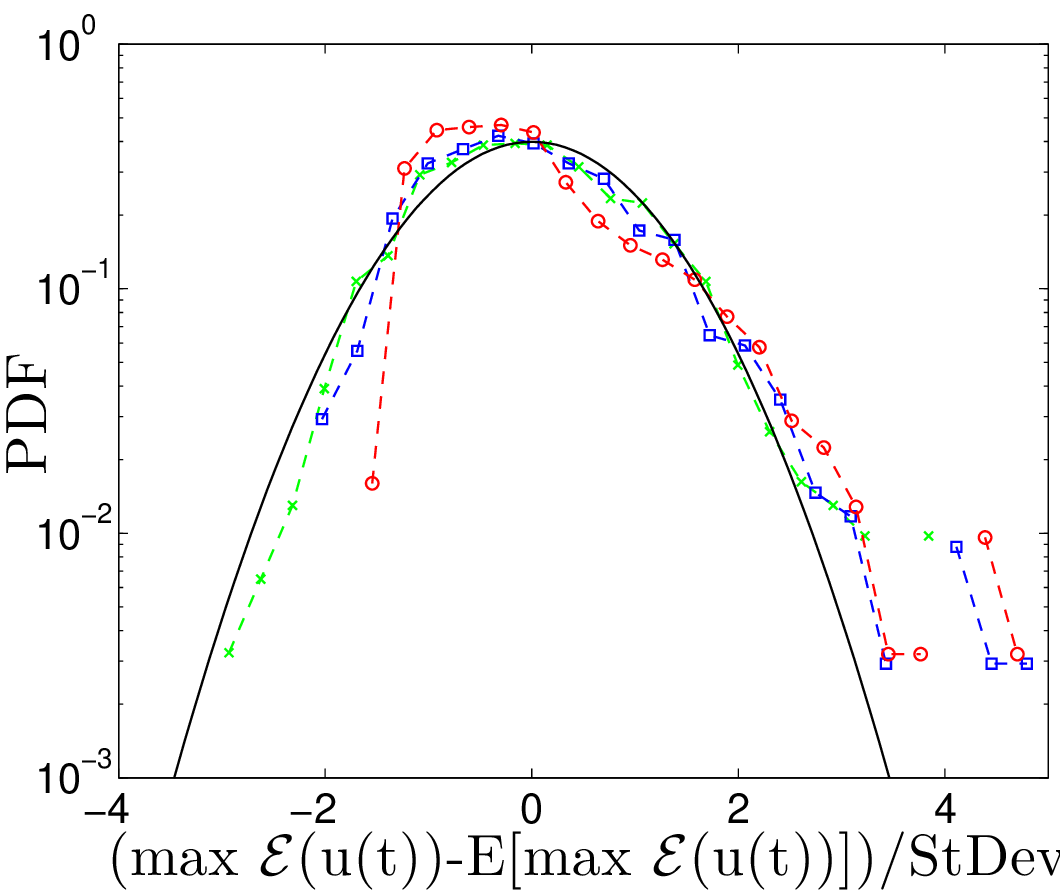}}\quad
\subfigure[]{\includegraphics[width=.475\textwidth]{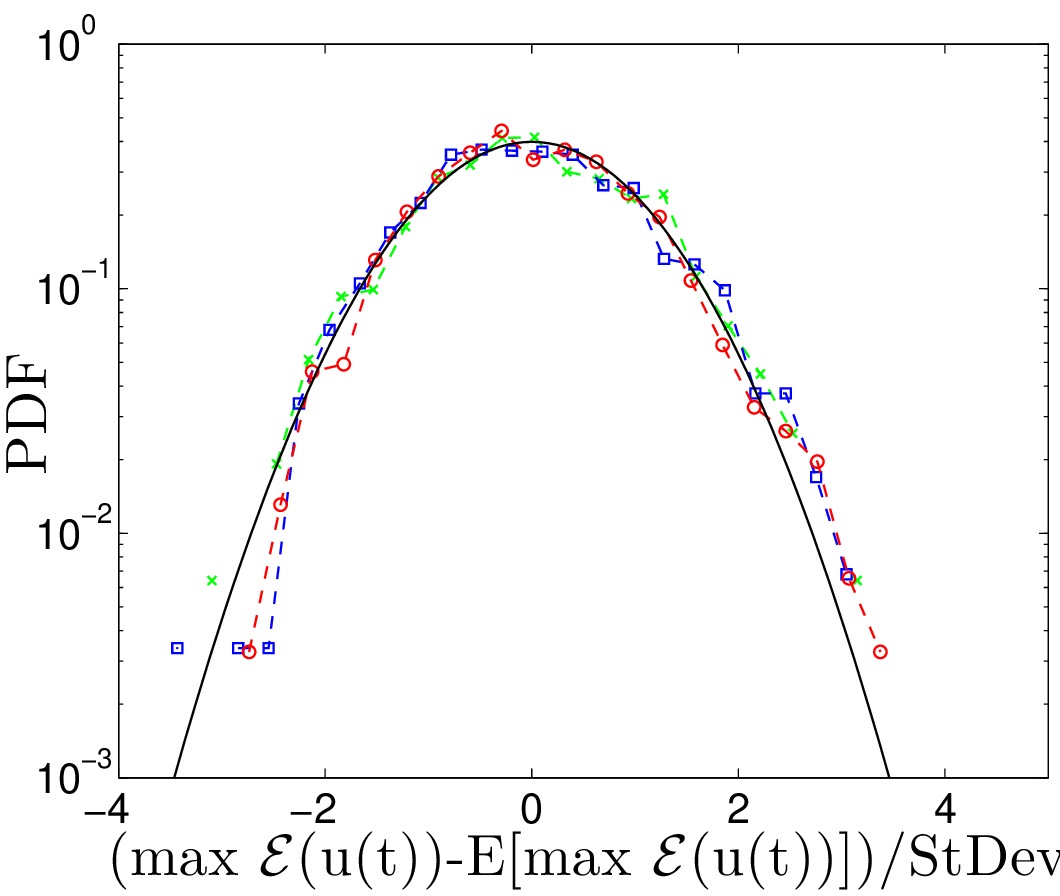}}
  \caption{{Normalized PDFs of the maximum enstrophy values
      $\max_{t \ge 0} \E(u(t,\omega))$ for {the cases with} the
      initial condition $\tg0$ with {$T=1$ and (a) $\Ecal_0 =
        10$, (b) $\Ecal_0 = 10^3$. The noise levels $\sigma^2$ are
        equal to} $10^{-2}$ {(green lines and crosses)}, $10^{-1}$
      {(blue lines and squares)} and $1$ {(red lines and circles)}. To
      obtain these plots, {$S=10^5$} samples were collected
      {in each case} and {sorted} into $30$ {equispaced} bins.
      The solid lines correspond to the standard Gaussian
      distributions.}}
  \label{fig:pdf}
\end{figure}

\subsection{Global Effect of Noise on Enstrophy Growth for Varying
  $\Ecal_0$}
\label{sec:effect_E0}

In this section we analyze how the diagnostic quantities 
\begin{subequations}
\label{eq:diag}
\begin{alignat}{2}
&\ebb[\Ecal({ u(T)})],&  \qquad & \Ecal(\ebb[{ u(T)}]), \label{eq:diagT} \\
\max_{t \in [0,T]} &\ebb[\Ecal({ u(t)})],&  \qquad  \max_{t \in [0,T]} & \Ecal(\ebb[{ u(t)}]) \label{eq:diagmax}
\end{alignat}
\end{subequations}
for some given $T$ depend on the initial enstrophy $\Ecal_0$ and
whether the presence of the stochastic excitation modifies the
power-law dependence of the quantities \eqref{eq:diagmax} on $\Ecal_0$
as compared to the deterministic case (cf.~Section
\ref{sec:deterministic}). We will do this in two cases, namely, when
for different values of the initial enstrophy $\E_0$ the noise level
$\sigma^2$ is fixed and when it is proportional to $\E_0$.
{Concerning the first case, Figures \ref{fig:fxtens}a and
  \ref{fig:fxtens}b} show the dependence of the quantities
\eqref{eq:diagT} and \eqref{eq:diagmax} with {$T=1$} on
$\Ecal_0$ for different fixed noise levels. The quantities
$\Ecal(\ebb[{ u(T)}])$ and $\ebb[\Ecal({ u(T)})]$ for
different time horizons $T$ are plotted as functions of $\Ecal_0$ for
small and large noise levels, respectively, in Figures
\ref{fig:smlsgmens} and \ref{fig:bigsgmens}.  These plots are
therefore the stochastic counterparts of Figure \ref{fig:Edet}
representing the deterministic case \cite{ap11a}.  We see that with a
fixed $T$ both $\Ecal(\ebb[{ u(T)}])$ and $\ebb[\Ecal({
  u(T)})]$ saturate at a level depending on the noise magnitude
$\sigma^2$ (Figure \ref{fig:fxtens}a). Analogous behavior is observed
for a fixed noise level and increasing time intervals in Figures
\ref{fig:smlsgmens} and \ref{fig:bigsgmens}, {from} which we can also
conclude that when we maximize the quantities $\Ecal(\ebb[{
  u(T)}])$ and $\ebb[\Ecal({ u(T)})]$ over all considered values
of $T$, then the resulting quantity will scale proportionally to
$\Ecal_0^{3/2}$, which is the same behavior as observed in the
deterministic case (Figure \ref{fig:Edet}). The process of maximizing
with respect to $T$ is represented schematically in Figures
\ref{fig:smlsgmens} and \ref{fig:bigsgmens} as ``envelopes'' of the
curves corresponding to different values of $T$.  {Regarding}
the behavior of the quantities \eqref{eq:diagmax}, for every noise
level we observe a transition from a noise-dominated behavior, where
$\max_{t \in [0,T]} \ebb[\Ecal({ u(t)})]$ does not increase with
$\Ecal_0$ when $\Ecal_0$ is small, to a nonlinearity-dominated regime
in which $\max_{t \in [0,T]} \ebb[\Ecal({ u(t)})]$ grows with
$\Ecal_0$ (Figure \ref{fig:fxtens}b). {In} the latter regime,
corresponding to large values of $\Ecal_0$ and whose lower bound is an
increasing function of the noise magnitude, we observe that for
sufficiently large $\Ecal_0$ the growth of the quantity $\max_{t \in
  [0,T]} \ebb[\Ecal({ u(t)})]$ in all cases approaches the growth
observed in the deterministic case \cite{ap11a}.
\begin{figure}
  \centering
\subfigure[]{\includegraphics[width=.45\textwidth]{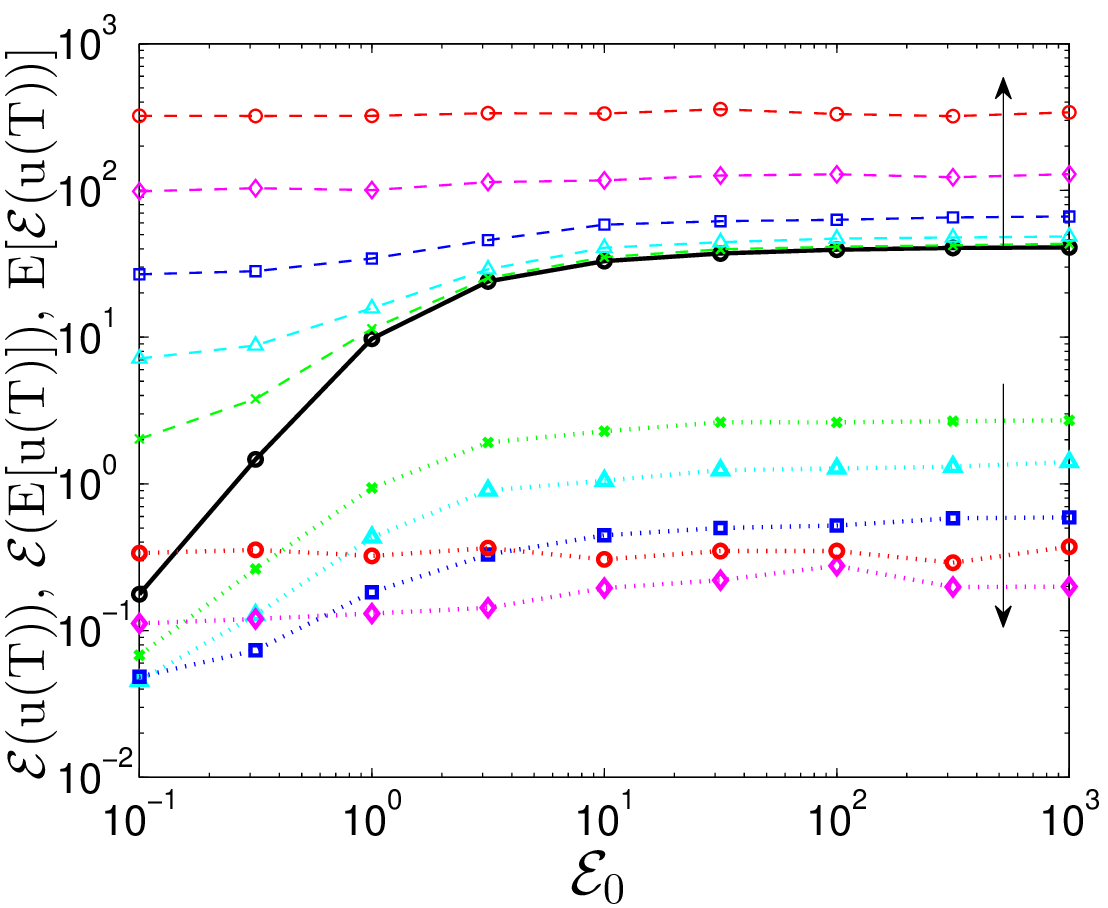}}\quad
\subfigure[]{\includegraphics[width=.45\textwidth]{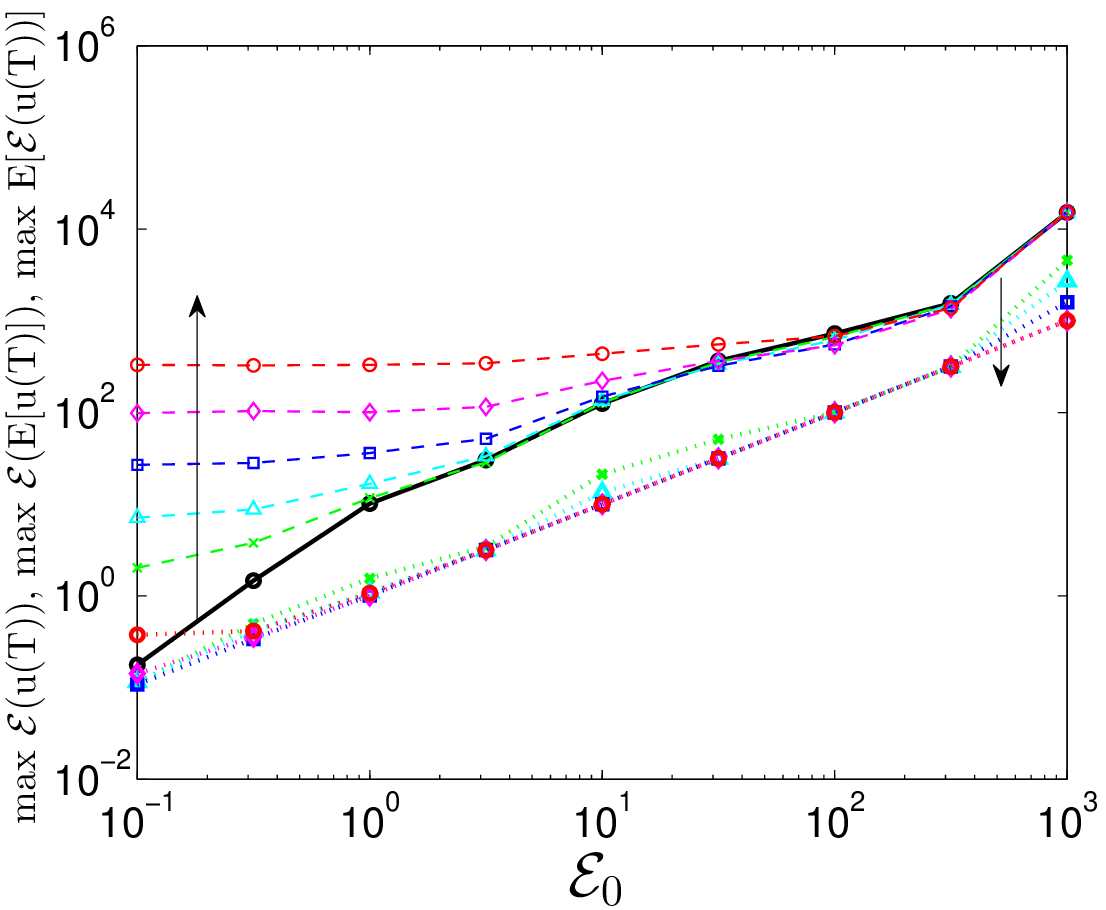}}
  \caption{(a) The values at {$T=1$} and (b) the maximum values
    attained in $[0,T]$ of the expected value of the enstrophy
    $\ebb[\Ecal({ u(t)})]$ (dashed lines), the enstrophy of the
    expected value of the solution $\Ecal(\ebb[{ u(t)}])$ (dotted
    lines) and the enstrophy $\Ecal(t)$ of the deterministic solution
    (thick solid line) as functions of the initial enstrophy $\Ecal_0$
    for the initial condition $\tg0$ with $\Ecal_0 = 10$, $T=1$ and
    different noise levels $\sigma^2$ in the range from $10^{-2}$ to
    $1$ (the direction of increase of $\sigma^2$ is indicated
    by arrows)}
    \label{fig:fxtens}
\end{figure}

\begin{figure}
  \centering
\subfigure[]{\includegraphics[width=.45\textwidth]{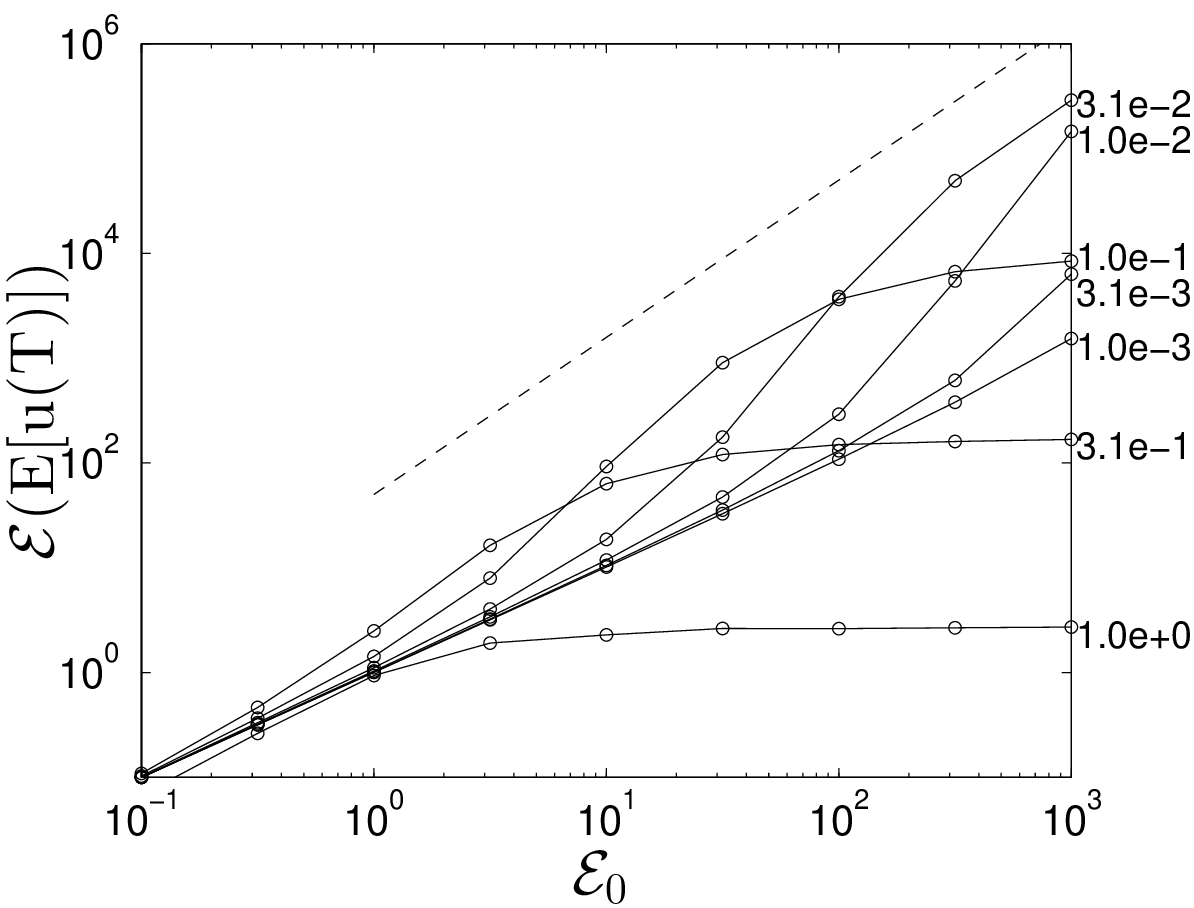}}\quad
\subfigure[]{\includegraphics[width=.42\textwidth]{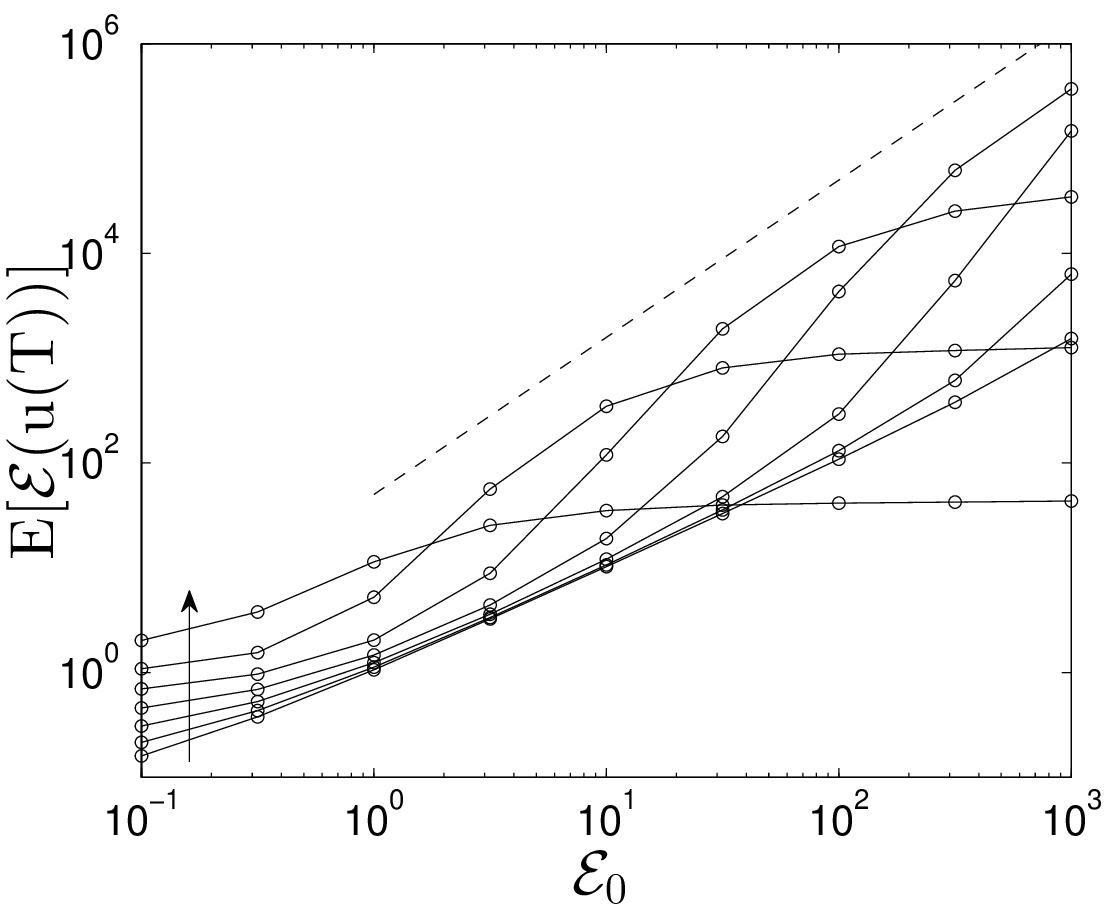}}
  \caption{[Small noise case: $\sigma^2 = 10^{-2}$] Dependence of (a)
    the enstrophy of the expected value of the solution
    $\Ecal(\ebb[{ u(T)}])$ and (b) the expected value of the
    enstrophy $\ebb[\Ecal({ u(T)})]$ on the initial enstrophy
    $\Ecal_0$ using the initial condition $\tg0$ with $T$ varying from
    $10^{-3}$ to $1$.  {In (a) the values of $T$ are marked
      near the right edge of the plot, whereas in (b) the direction of
      increasing $T$ is indicated with an arrow.}  The {dashed} lines
    correspond to the power law $C \, \Ecal_0^{3/2}$.}
\label{fig:smlsgmens}
\bigskip\bigskip
\subfigure[]{\includegraphics[width=.45\textwidth]{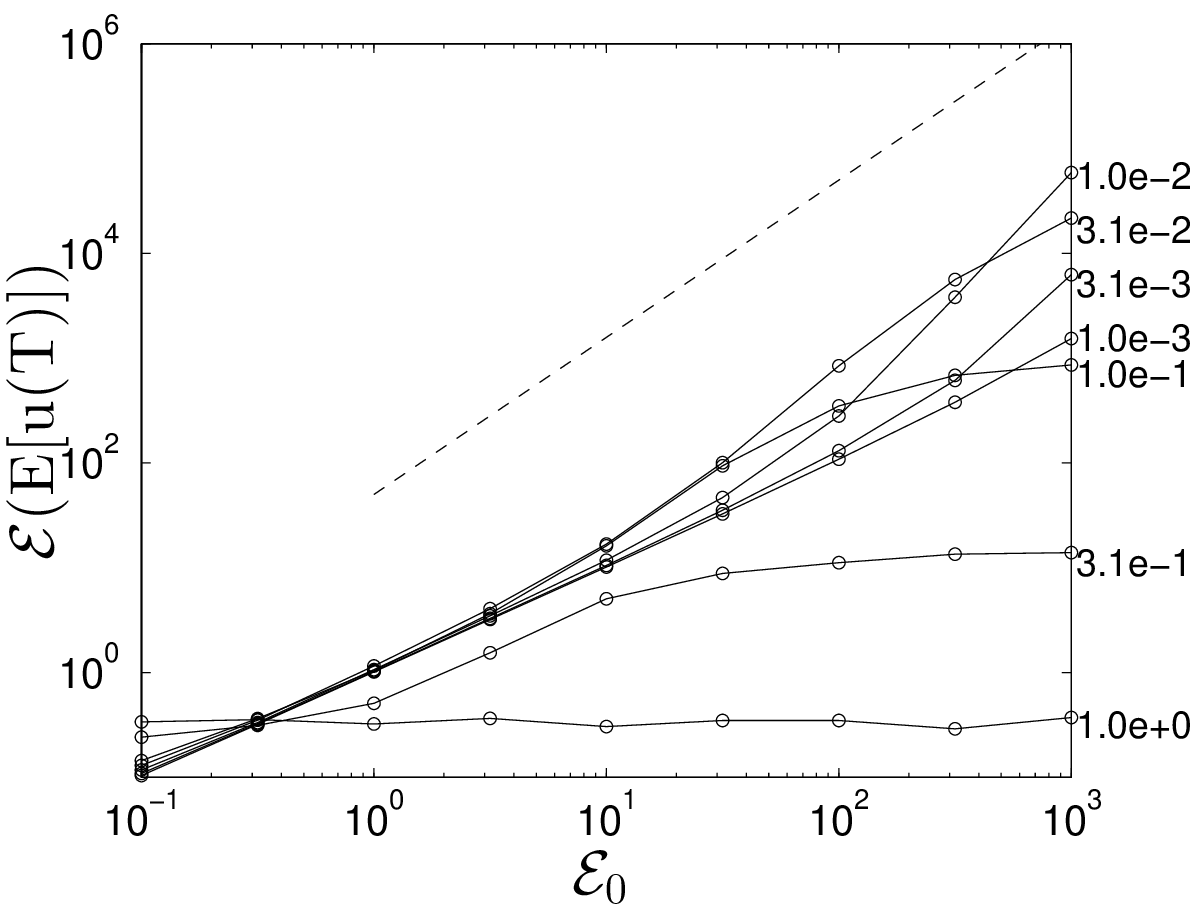}}\quad
\subfigure[]{\includegraphics[width=.42\textwidth]{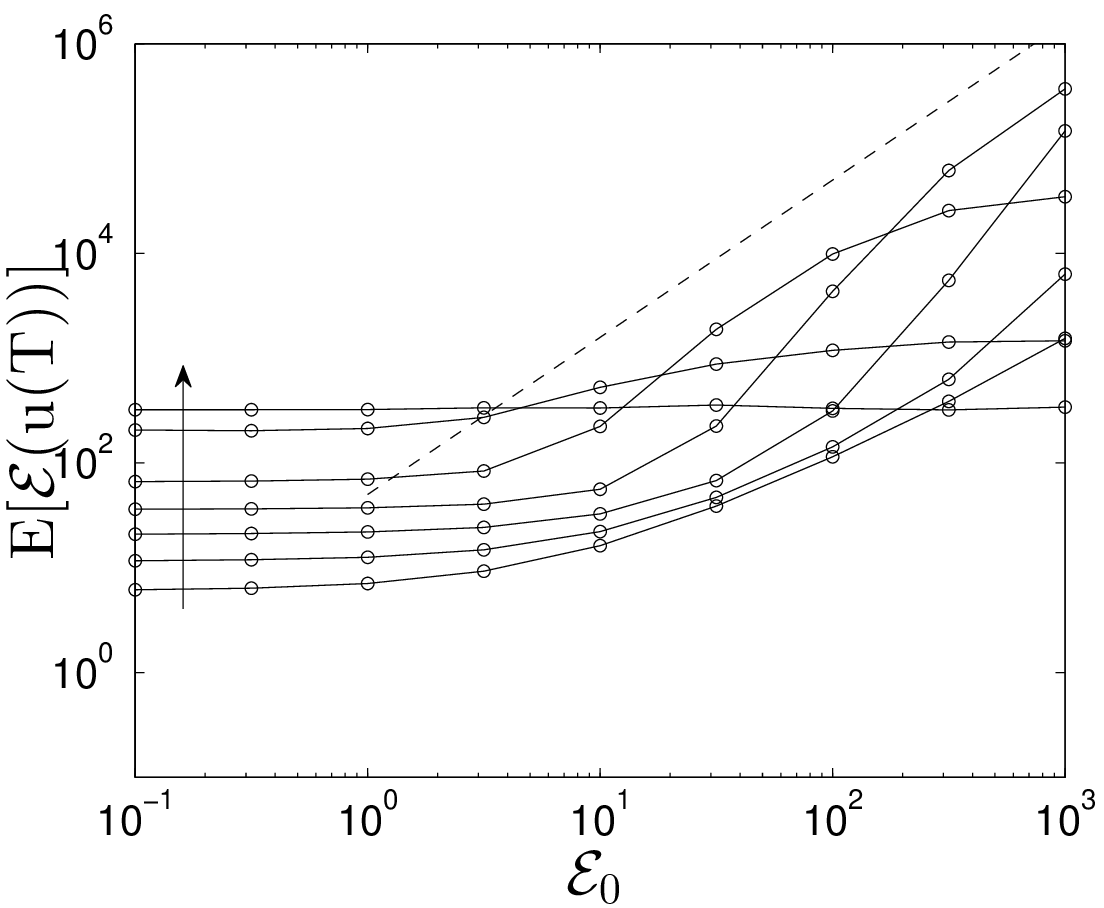}}
  \caption{[Large noise case: $\sigma^2=1$] (see previous figure for
    details).}
\label{fig:bigsgmens}
\end{figure}

Since the results presented above show no evidence of the effect of
noise on the dependence of the quantities \eqref{eq:diag} on
$\Ecal_0$ when $\Ecal_0$ grows while the noise magnitude stays fixed,
to close this section we consider the case in which the noise
magnitude is proportional to $\Ecal_0$, i.e.,
\begin{equation}
\sigma^2 = C_{\sigma} \, \Ecal_0,
\label{eq:sigE0}
\end{equation}
for a range of different constants $C_{\sigma}$. The quantities
\eqref{eq:diagmax} obtained in this way are shown in Figures
\ref{fig:diagonal}a and \ref{fig:diagonal}b. As regards the dependence
of the quantity $\max_{t \in [0,T]} \Ecal(\ebb[{ u(t)}])$ on
$\Ecal_0$, in Figure \ref{fig:diagonal}a we observe a superlinear
growth which is however slower than $\Ecal_0^{3/2}$ characterizing the
deterministic case (in fact, from the data it is not entirely obvious
if this dependence is strictly in the form of a power law). Concerning
the quantity $\max_{t \in [0,T]} \ebb[\Ecal({ u(t)})]$, Figure
\ref{fig:diagonal}b indicates that while for small $\Ecal_0$ it is
larger than $\max_{t \ge 0} \Ecal(t)$ obtained in the deterministic
case, in the limit of $\Ecal_0 \rightarrow \infty$ it reveals the same
growth as in the deterministic case, that is, proportional to
$\Ecal_0^{3/2}$ with approximately the same constant prefactor.
\begin{figure}
  \centering
\subfigure[]{\includegraphics[width=.45\textwidth]{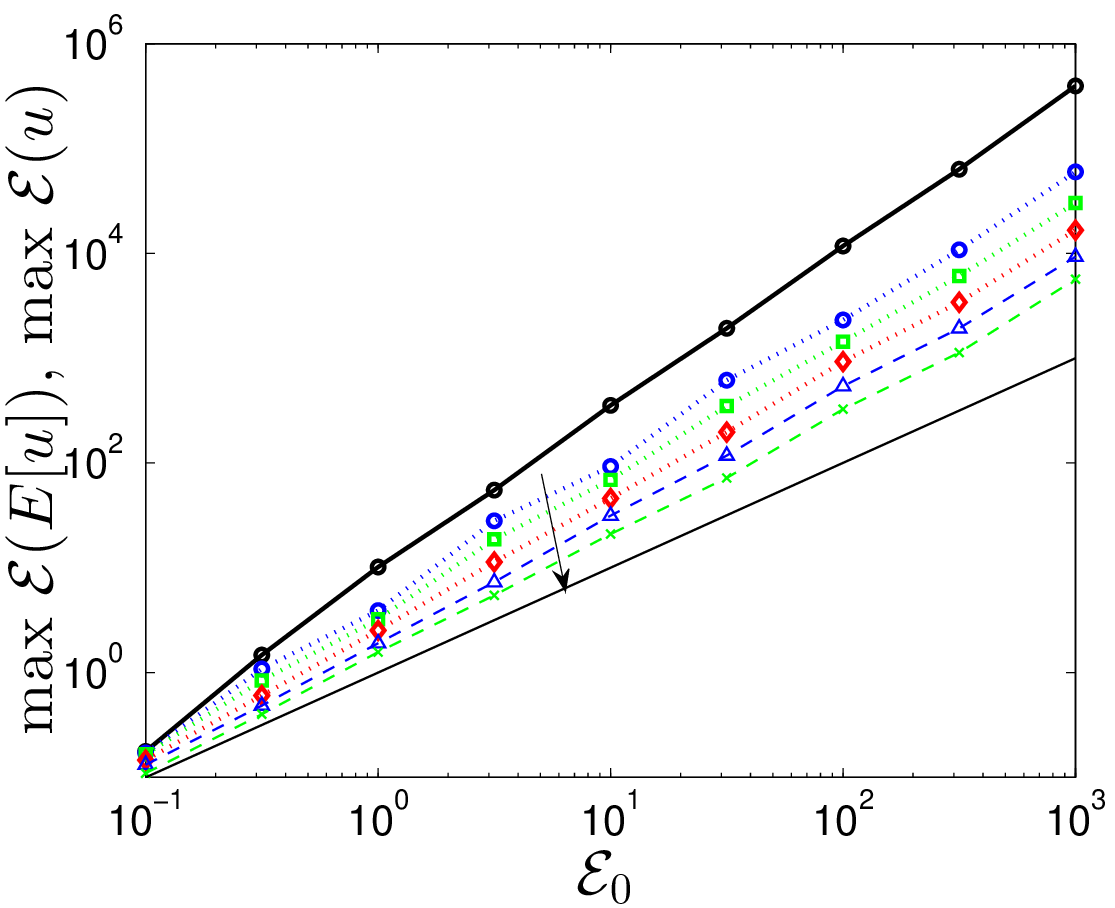}}\quad
\subfigure[]{\includegraphics[width=.45\textwidth]{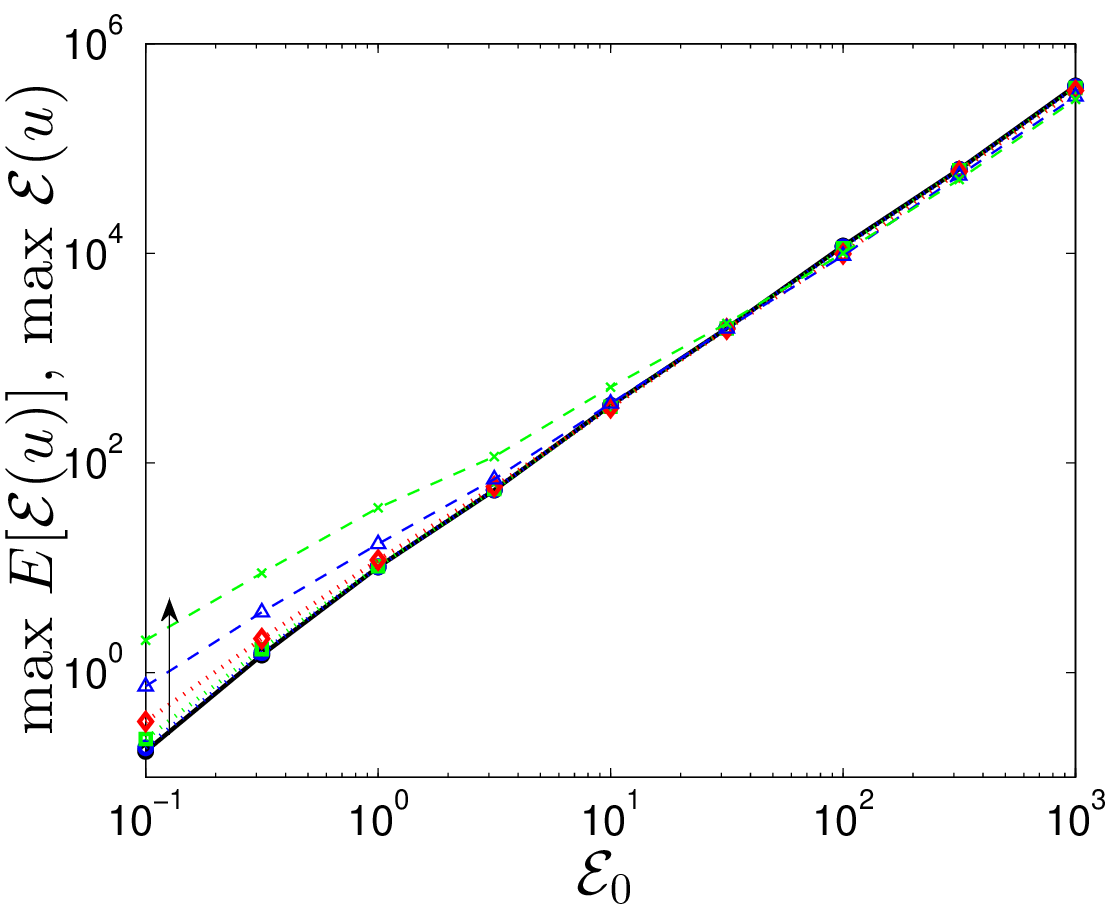}}
  \caption{Dependence of (a) the maximum enstrophy of the expected
    value of the solution $\max_{t\in[0,T]} \Ecal(\ebb[{ u(t)}])$
    and (b) the maximum expected value of the enstrophy
    $\max_{t\in[0,T]} \ebb[\Ecal({ u(t)})]$ on the initial
    enstrophy $\Ecal_0$ using the initial conditions $\tg0$ and with
    noise magnitudes proportional to $\Ecal_0$, cf.~\eqref{eq:sigE0},
    with $C_{\sigma}$ {in the range from $10^{-3}$ to $10^{-1}$}
    (arrow indicate the direction of increase of $C_{\sigma}$). The
    parameter $T$ is chosen to maximize $\max_{t\in[0,T]}
    \Ecal(\ebb[{ u(t)}])$ in (a) and $\max_{t\in[0,T]}
    \ebb[\Ecal({ u(t)})]$ in (b). The thick black solid line
    corresponds to the quantity $\max_{t \in [0,T]} \Ecal(t)$ obtained
    in the deterministic case, whereas the thin black solid line in
    (a) represents the power law $\Ecal_0^1$.}
\label{fig:diagonal}
\end{figure}

\section{Discussion and Conclusions}
\label{sec:final}

The goal of this study was to test whether a stochastic excitation
applied to Burgers equation can affect the maximum growth of enstrophy
as a function of the initial enstrophy $\Ecal_0$ observed in the
deterministic case \cite{ap11a}. In the context of
hydrodynamic models based on the Navier-Stokes equation, the enstrophy
is a convenient indicator of the regularity of solutions and its
growth is inherently related to the problem of finite-time singularity
formation \cite{d09}. In the stochastic problem considered here, there
are two relevant quantities related to the enstrophy, namely, the
expected value of the enstrophy $\ebb[\Ecal(u(t))]$ and the enstrophy
of the expected value of the solution $\Ecal(\ebb[u(t)])$. They
  are related to each other via Jensen's inequality
  \eqref{eq:jensen}. In the set-up of our problem we allowed for the
most ``aggressive'' form of the stochastic excitation which still
ensures that the two quantities are well defined (cf.~Section
\ref{sec:noise}). The numerical discretization was carefully designed
based on the Monte Carlo sampling.

The effect of the noise was found to depend on the relation between
its magnitude {$\sigma^2$} and the ``size'' of the initial data as
measured by the initial enstrophy $\Ecal_0$. When the noise magnitude
is large, the stochastic excitation obscures the intrinsic dynamics
and any dependence of the diagnostic quantities \eqref{eq:diag} on
$\Ecal_0$ is lost.  Therefore, the relevant regime is when the noise
magnitude is ``modest'' relative to the initial enstrophy $\Ecal_0$,
so that the stochastic excitation can be regarded as a
``perturbation'' of the deterministic dynamics. We observe that the
two quantities $\ebb[\Ecal(u(t))]$ and $\Ecal(\ebb[u(t)])$ provide,
respectively, upper and lower bounds on the enstrophy $\Ecal(t)$ in
the deterministic case, cf.~\eqref{eq:EEE}, with the bounds becoming
tighter as the noise magnitude vanishes (Figure \ref{fig:incsgm}).
{The fact that the deterministic enstrophy $\E(t)$ is
  ``bracketed'' by $\ebb[\Ecal(u(t))]$ and $\Ecal(\ebb[u(t)])$ appears
  to be a new, though not entirely unexpected, finding.}  The latter
case, with the enstrophy of the expected value of the solution
$\Ecal(\ebb[u(t)])$ being lower than the deterministic enstrophy
$\Ecal(t)$, can be therefore interpreted in terms of the stochastic
excitation having the effect of an increased dissipation of the
expected value of the solution.

{The non-Gaussian PDFs of the normalized maximum enstrophy $\max_{t
    \ge 0} \E({u(t,\omega)})$ in Figures \ref{fig:pdf}(a,b) indicate
  the likelihood of events when larger-than-average enstrophy maxima
  are achieved, although this property becomes less pronounced as the
  enstrophy $\E_0$ of the initial condition grows. This can be
  interpreted to mean that as the magnitude of the nonlinear effects
  increases, the transient evolution becomes less susceptible to
  stochastic excitation.} {We note that non-Gaussian PDFs of solution
  derivatives $\partial_x u$ in stochastic Burgers flows were also
  reported and analyzed in {\cite{cy95a,cy95b,ztg97,gk93}} (since in
  those studies the PDFs were computed for a different quantity, the
  actual shapes of the distributions {and their dependence on
    parameters} were different).}

As regards the expected value of the enstrophy, we observed in Figure
\ref{fig:diagonal}a that in the limit $\Ecal_0 \rightarrow \infty$ the
quantity $\max_{t \ge 0} \ebb[\Ecal(u(t))]$ exhibits the same
dependence on $\Ecal_0$ as in the deterministic case, i.e., it remains
proportional to $\Ecal_0^{3/2}$, even for the noise magnitude
increasing proportionally to $\Ecal_0$.  Thus, this demonstrates that
the stochastic excitation {\em does not} {damp} the maximum
growth of enstrophy as a function of the initial enstrophy $\Ecal_0$.
{This observation is {further} reinforced by {the
    PDFs of $\max_{t \ge 0} \E({u(t,\omega)})$ shown in Figures
    \ref{fig:pdf}(a,b) which are skewed towards values larger than
    $\ebb[{\max_{t \ge 0}\Ecal(u(t))}]$, but approach the Gaussian
    distribution as $\E_0$ increases}. In the light of the findings
  reported in \cite{ar09,ar10}, where it was shown that a certain
  stochastic excitation can regularize the inviscid Burgers equation,
  our result does not appear entirely obvious. It can be however
  interpreted as a consequence of the robustness of the
  shock-formation process which is not disturbed by stochastic
  excitation. If these insights could be extrapolated to the 3D case,
  one could expect that noise would be less likely to regularize the
  3D Navier-Stokes system than the corresponding Euler system.}

We note that if we rescale the magnitude of the solution $u$ as
  $u_a=a \, u$ for some $a>0$, then the stochastic Burgers equation
  \eqref{eq:burgersA} will be left invariant if we simultaneously
  rescale the time, viscosity and the forcing term as $t_a=t/a$,
  $\nu_a=a \, \nu$ and $\zeta_a = a^2 \, \zeta$. Therefore, the limit
  $\E_0\rightarrow\infty$ (while keeping $\nu$ fixed) considered in
  the present study is equivalent to the limit $\nu \rightarrow 0$
  (while keeping initial data fixed) which was investigated in other
  studies \cite{s92,saf92,ekms00}. In particular, it was shown in
  \cite{ekms00} that inclusion of additive noise in the inviscid
  Burgers equation significantly increases the number of shocks. This
  results is however not inconsistent with our findings, since it
  corresponds to stochastic forcing with a {\em finite} magnitude,
  whereas for the problem set-up considered here the limit $\nu
  \rightarrow 0$ would imply vanishing magnitude (at the quadratic
  rate) of the forcing term.

A number of related questions remain open. First of all, in the
present study we numerically solved the stochastic Burgers equation
\eqref{eq:burgers} using the extreme initial data $\tg0$ which was
found in \cite{ap11a} by solving a deterministic variational
optimization problem. It is however possible that by solving a
corresponding {\em stochastic} optimization problem one might obtain
initial data $g$ leading to an even larger growth of enstrophy in
finite time. While such problems are harder to solve than the
deterministic one, they are in principle amenable to solution using
stochastic programming methods \cite{r06}. {We add that this
  approach would be distinct from the ``instanton'' formulation
  \cite{mkv16,bfkl97,ggs15} which due to the saddle-point
  approximation is effectively equivalent to solution of a
  deterministic optimization problem.}  In a similar spirit, it is
equally interesting to obtain {\em rigorous} estimates on $d\Ecal /
dt$ and $\max_{t \ge 0} \Ecal(t)$ in the stochastic setting in terms
of {$\Ecal_0$} and the properties of noise, thereby generalizing the
bounds available for the deterministic case \cite{ld08,ap11a}.
{As regards effects of viscous dissipation, it is well known
  \cite{knv06} that the {\em fractional} Burgers equation is no longer
  globally well posed when the fractional dissipation exponent $\alpha
  < 1/2$. It would be therefore interesting to see whether the
  finite-time blow-up known to occur in this supercritical regime can
  be mollified by noise.}  Similar questions concerning the interplay
between the stochastic excitation and extreme behavior, including
possible singularity formation, also arise in the context of the
two-dimensional and three-dimensional Navier-Stokes and Euler
equations. Addressing at least some of these issues is one of the
goals of the ongoing research program mentioned in Introduction.

\section*{Acknowledgements}

The authors would like to thank Dr.~Diego Ayala for his help
  with computing the optimal initial data $\tg0$ \cite{ap11a}. This
research was supported by NSERC (Canada) and FCT Doctoral Grant
(Portugal).

\providecommand{\href}[2]{#2}
\providecommand{\arxiv}[1]{\href{http://arxiv.org/abs/#1}{arXiv:#1}}
\providecommand{\url}[1]{\texttt{#1}}
\providecommand{\urlprefix}{URL }


\begin{thebibliography}{10}

\bibitem{af05}
\newblock R.~A. Adams and J.~F. Fournier,
\newblock \emph{{Sobolev} Spaces},
\newblock Elsevier, 2005.

\bibitem{AG06}
\newblock A.~Alabert and I.~Gy\"ongy,
\newblock {On numerical approximation of stochastic Burgers' equation},
\newblock in \emph{From stochastic calculus to mathematical finance},
\newblock Springer, 2006,
\newblock 1--15.

\bibitem{ar09}
\newblock S.~Albeverio and O.~Rozanova,
\newblock The non-viscous {Burgers} equation associated with random position in
  coordinate space: a threshold for blow up behaviour,
\newblock \emph{Mathematical Models and Methods in Applied Sciences},
  \textbf{19} (2009), 749--767,
\newblock
  \urlprefix\url{http://www.worldscientific.com/doi/abs/10.1142/S0218202509003%
607}.

\bibitem{ar10}
\newblock S.~Albeverio and O.~Rozanova,
\newblock Suppression of unbounded gradients in an {SDE} associated with the
  {Burgers} equation,
\newblock \emph{Proceedings of the American Mathematical Society}, \textbf{138}
  (2010), 241--251,
\newblock \urlprefix\url{http://www.jstor.org/stable/40590613}.

\bibitem{ap11a}
\newblock D.~Ayala and B.~Protas,
\newblock On maximum enstrophy growth in a hydrodynamic system,
\newblock \emph{Physica D}, \textbf{240} (2011), 1553--1563.

\bibitem{ap13a}
\newblock D.~Ayala and B.~Protas,
\newblock Maximum palinstrophy growth in {2D} incompressible flows,
\newblock \emph{Journal of Fluid Mechanics}, \textbf{742} (2014), 340--367.

\bibitem{ap13b}
\newblock D.~Ayala and B.~Protas,
\newblock Vortices, maximum growth and the problem of finite-time singularity
  formation,
\newblock \emph{Fluid Dynamics Research}, \textbf{46} (2014), 031404.

\bibitem{ap16}
\newblock D.~Ayala and B.~Protas,
\newblock Extreme vortex states and the growth of enstrophy in {3D}
  incompressible flows,
\newblock \emph{Journal of Fluid Mechanics}, \textbf{818} (2017), 772--806.

\bibitem{bfkl97}
\newblock E.~Balkovsky, G.~Falkovich, I.~Kolokolov and V.~Lebedev,
\newblock {Intermittency of Burgers' Turbulence},
\newblock \emph{Phys. Rev. Lett.}, \textbf{78} (1997), 1452--1455,
\newblock \urlprefix\url{http://link.aps.org/doi/10.1103/PhysRevLett.78.1452}.

\bibitem{bk07}
\newblock J.~Bec and K.~Khanin,
\newblock Burgers turbulence,
\newblock \emph{Physics Reports}, \textbf{447} (2007), 1--66,
\newblock
  \urlprefix\url{http://www.sciencedirect.com/science/article/pii/S03701573070%
01457}.

\bibitem{BCJL94}
\newblock L.~Bertini, N.~Cancrini and G.~Jona-Lasinio,
\newblock {The stochastic Burgers equation},
\newblock \emph{Communications in Mathematical Physics}, \textbf{165} (1994),
  211--232.

\bibitem{BJ13}
\newblock D.~Blomker and A.~Jentzen,
\newblock {Galerkin approximations for the stochastic Burgers equation},
\newblock \emph{SIAM Journal on Numerical Analysis}, \textbf{51} (2013),
  694--715.

\bibitem{b14}
\newblock A.~Boritchev,
\newblock Decaying turbulence in the generalised {Burgers} equation,
\newblock \emph{Archive for Rational Mechanics and Analysis}, \textbf{214}
  (2014), 331--357,
\newblock \urlprefix\url{http://dx.doi.org/10.1007/s00205-014-0766-5}.

\bibitem{canuto:SpecMthd}
\newblock C.~Canuto, A.~Quarteroni, Y.~Hussaini and T.~A. Zang,
\newblock \emph{Spectral Methods},
\newblock Scientific Computation, Springer, 2006.

\bibitem{cy95a}
\newblock A.~Chekhlov and V.~Yakhot,
\newblock Kolmogorov turbulence in a random-force-driven {Burgers} equation,
\newblock \emph{Phys. Rev. E}, \textbf{51} (1995), R2739--R2742,
\newblock \urlprefix\url{http://link.aps.org/doi/10.1103/PhysRevE.51.R2739}.

\bibitem{cy95b}
\newblock A.~Chekhlov and V.~Yakhot,
\newblock Kolmogorov turbulence in a random-force-driven {Burgers} equation:
  Anomalous scaling and probability density functions,
\newblock \emph{Phys. Rev. E}, \textbf{52} (1995), 5681--5684,
\newblock \urlprefix\url{http://link.aps.org/doi/10.1103/PhysRevE.52.5681}.

\bibitem{davidson:turbulence}
\newblock P.~A. Davidson,
\newblock \emph{Turbulence. An introduction for scientists and engineers},
\newblock Oxford University Press, 2004.

\bibitem{ddm02a}
\newblock A.~Debussche and L.~D. Menza,
\newblock {Numerical simulation of focusing stochastic nonlinear Schr\"odinger
  equations},
\newblock \emph{Physica D}, \textbf{162} (2002), 131--154.

\bibitem{d09}
\newblock C.~R. Doering,
\newblock The {3D Navier-Stokes} problem,
\newblock \emph{Annual Review of Fluid Mechanics}, \textbf{41} (2009),
  109--128.

\bibitem{ekms00}
\newblock W.~E, K.~Khanin, A.~Mazel and Y.~Sinai,
\newblock Invariant measures for burgers equation with stochastic forcing,
\newblock \emph{Annals of Mathematics}, \textbf{151} (2000), 877--960,
\newblock \urlprefix\url{http://www.jstor.org/stable/121126}.

\bibitem{f00}
\newblock C.~L. Fefferman,
\newblock Existence and smoothness of the {Navier-Stokes} equation,
\newblock \urlprefix\url{http://www.claymath.org/sites/default/files/navierstokes.pdf}, 2000,
\newblock {Clay Millennium Prize Problem Description}.

\bibitem{f15}
\newblock F.~Flandoli,
\newblock \emph{{Random Perturbation of PDEs and Fluid Dynamic Models}},
\newblock Lecture Notes in Mathematics, Springer, 2015.

\bibitem{f15b}
\newblock F.~Flandoli,
\newblock \emph{{Stochastic Analysis: A Series of Lectures (Eds. R.C. Dalang,
  M. Dozzi, F. Flandoli and F. Russo)}}, chapter {A Stochastic View over the
  Open Problem of Well-posedness for the 3D Navier-Stokes Equations}, 221--246,
\newblock Birkh\"auser, 2015.

\bibitem{gbk08}
\newblock J.~D. Gibbon, M.~Bustamante and R.~M. Kerr,
\newblock The three--dimensional {Euler} equations: singular or non--singular?,
\newblock \emph{Nonlinearity}, \textbf{21} (2008), 123--129.

\bibitem{gk93}
\newblock T.~Gotoh and R.~H. Kraichnan,
\newblock Statistics of decaying {Burgers} turbulence,
\newblock \emph{Physics of Fluids A}, \textbf{5} (1993), 445--457,
\newblock
  \urlprefix\url{http://scitation.aip.org/content/aip/journal/pofa/5/2/10.1063%
/1.858868}.

\bibitem{ggs15}
\newblock T.~Grafke, R.~Grauer and T.~Sch{\"a}fer,
\newblock The instanton method and its numerical implementation in fluid
  mechanics,
\newblock \emph{Journal of Physics A: Mathematical and Theoretical},
  \textbf{48} (2015), 333001,
\newblock \urlprefix\url{http://stacks.iop.org/1751-8121/48/i=33/a=333001}.

\bibitem{g98}
\newblock I.~Gy\"ongy,
\newblock Existence and uniqueness results for semilinear stochastic partial
  differential equations,
\newblock \emph{Stochastic Processes and their Applications}, \textbf{73}
  (1998), 271--299.

\bibitem{GN99}
\newblock I.~Gy\"ongy and D.~Nualart,
\newblock {On the stochastic Burgers' equation in the real line},
\newblock \emph{The Annals of Probability}, \textbf{27} (1999), 782--802.

\bibitem{HM15}
\newblock M.~Hairer and K.~Matetski,
\newblock {Optimal rate of convergence for stochastic Burgers-type equations},
  2015,
\newblock ArXiv:1504:05134.

\bibitem{h14a}
\newblock M.~Hairer,
\newblock {Solving the KPZ equation},
\newblock \emph{Annals of Mathematics}, \textbf{178} (2014), 559--664.

\bibitem{Kim2007arfm}
\newblock J.~Kim and T.~Bewley,
\newblock A linear systems approach to flow control,
\newblock \emph{Ann.\ Rev.\ Fluid Mech.}, \textbf{39} (2007), 383--417.

\bibitem{knv06}
\newblock A.~Kiselev, F.~Nazarov and A.~Volberg,
\newblock Global well-posedness for the critical {2D} dissipative
  quasi-geostrophic equation,
\newblock \emph{Invent. math.}, \textbf{167} (2006), 445--453.

\bibitem{ks15a}
\newblock C.~Klein and J.-C. Saut,
\newblock {A numerical approach to blow-up issues for dispersive perturbations
  of Burgers' equation},
\newblock \emph{Physica D}, \textbf{295--296} (2015), 46--65.

\bibitem{kl04}
\newblock H.~Kreiss and J.~Lorenz,
\newblock \emph{Initial-Boundary Value Problems and the {Navier-Stokes}
  Equations}, vol.~47 of Classics in Applied Mathematics,
\newblock SIAM, 2004.

\bibitem{ks12}
\newblock S.~Kuksin and A.~Shirikyan,
\newblock \emph{Mathematics of two-dimensional turbulence},
\newblock Cambridge University Press, 2012.

\bibitem{lps14}
\newblock G.~J. Lord, C.~E. Powell and T.~Shardlow,
\newblock \emph{An Introduction to Computational Stochastic PDEs},
\newblock Cambridge University Press, 2014.

\bibitem{ld08}
\newblock L.~Lu and C.~R. Doering,
\newblock Limits on enstrophy growth for solutions of the three-dimensional
  {Navier--Stokes} equations,
\newblock \emph{Indiana University Mathematics Journal}, \textbf{57} (2008),
  2693--2727.

\bibitem{mkv16}
\newblock B.~Meerson, E.~Katzav and A.~Vilenkin,
\newblock Large deviations of surface height in the {Kardar-Parisi-Zhang}
  equation,
\newblock \emph{Phys. Rev. Lett.}, \textbf{116} (2016), 070601,
\newblock
  \urlprefix\url{http://link.aps.org/doi/10.1103/PhysRevLett.116.070601}.

\bibitem{p12b}
\newblock D.~Pelinovsky,
\newblock Enstrophy growth in the viscous {Burgers} equation,
\newblock \emph{Dynamics of Partial Differential Equations}, \textbf{9} (2012),
  305--340.

\bibitem{p12}
\newblock D.~Pelinovsky,
\newblock Sharp bounds on enstrophy growth in the viscous {Burgers} equation,
\newblock \emph{Proceedings of Royal Society A}, \textbf{468} (2012),
  3636--3648.

\bibitem{Pope2000book}
\newblock S.~Pope,
\newblock \emph{Turbulent Flows},
\newblock 1st edition,
\newblock Cambridge University Press, Cambridge, UK, 2000.

\bibitem{DPDT94}
\newblock G.~D. Prato, A.~Debussche and R.~Temam,
\newblock {Stochastic Burgers' equation},
\newblock \emph{Nonlinear Differential Equations and Applications}, \textbf{1}
  (1994), 289--402.

\bibitem{r06}
\newblock A.~Ruszczy\'nski,
\newblock \emph{Nonlinear Optimization},
\newblock Princeton University Press, 2006.

\bibitem{saf92}
\newblock Z.-S. She, E.~Aurell and U.~Frisch,
\newblock The inviscid {Burgers} equation with initial data of brownian type,
\newblock \emph{Communications in Mathematical Physics}, \textbf{148} (1992),
  623--641,
\newblock \urlprefix\url{http://dx.doi.org/10.1007/BF02096551}.

\bibitem{s92}
\newblock Y.~G. Sinai,
\newblock Statistics of shocks in solutions of inviscid {Burgers} equation,
\newblock \emph{Communications in Mathematical Physics}, \textbf{148} (1992),
  601--621,
\newblock \urlprefix\url{http://dx.doi.org/10.1007/BF02096550}.

\bibitem{ztg97}
\newblock O.~Zikanov, A.~Thess and R.~Grauer,
\newblock Statistics of turbulence in a generalized random-force-driven
  {Burgers} equation,
\newblock \emph{Physics of Fluids}, \textbf{9} (1997), 1362--1367,
\newblock
  \urlprefix\url{http://scitation.aip.org/content/aip/journal/pof2/9/5/10.1063%
/1.869250}.

\end{thebibliography}
\end{document}